\documentclass[%
 reprint,
 amsmath,amssymb,
 aps,
]{revtex4-1}

\usepackage{graphicx}%
\usepackage{dcolumn}%
\usepackage{bm}%
\usepackage{hyperref}%

\usepackage{siunitx}
\usepackage{float}
\usepackage{color}
\usepackage{amssymb}
\usepackage{wasysym}
\usepackage{multirow}
\usepackage[normalem]{ulem}%
\usepackage{upgreek}
\usepackage{braket}

\newcommand{\omegap}{\omega_\text{p}}
\newcommand{\omegac}{\omega_\text{c}}
\newcommand{\omegamu}{\omega_{\upmu}}
\newcommand{\omegam}{\omega_{\textrm{m}}}

\newcommand{\kappae}{\kappa_{\textrm{e}}}
\newcommand{\kappai}{\kappa_{\textrm{i}}}

\newcommand{\aout}{a_{\textrm{out}}}

\newcommand{\etae}{\eta_{\textrm{e}}}
\newcommand{\etao}{\eta_{\textrm{o}}}
\newcommand{\etam}{\eta_{\textrm{m}}}

\newcommand{\Cab}{C_{\textrm{ab}}}
\newcommand{\Cbc}{C_{\textrm{bc}}}

\newcommand{\gammae}{\gamma_{\textrm{e}}}
\newcommand{\gammamu}{\gamma_{\upmu}}

\newcommand{\nc}{n_{\textrm{c}}}

\newcommand{\nphon}{n_{\textrm{phon}}}

\newcommand{\alphap}{\alpha_{\textrm{p}}}

\newcommand{\Pin}{P_{\textrm{in}}}
\newcommand{\Pmu}{P_{\upmu}}

\begin{document}

\preprint{APS/123-QED}

\title{Efficient bidirectional piezo-optomechanical transduction between microwave and optical frequency}%

\author{Wentao Jiang}
\email{wentao@stanford.edu}
\affiliation{Department of Applied Physics and Ginzton Laboratory, Stanford University, 348 Via Pueblo Mall, Stanford, California 94305, USA}
\author{Christopher J. Sarabalis}
\affiliation{Department of Applied Physics and Ginzton Laboratory, Stanford University, 348 Via Pueblo Mall, Stanford, California 94305, USA}
\author{Yanni D. Dahmani}
\affiliation{Department of Applied Physics and Ginzton Laboratory, Stanford University, 348 Via Pueblo Mall, Stanford, California 94305, USA}
\author{Rishi N. Patel}
\affiliation{Department of Applied Physics and Ginzton Laboratory, Stanford University, 348 Via Pueblo Mall, Stanford, California 94305, USA}
\author{Felix M. Mayor}
\affiliation{Department of Applied Physics and Ginzton Laboratory, Stanford University, 348 Via Pueblo Mall, Stanford, California 94305, USA}
\author{Timothy P. McKenna}
\affiliation{Department of Applied Physics and Ginzton Laboratory, Stanford University, 348 Via Pueblo Mall, Stanford, California 94305, USA}
\author{Rapha\"el Van Laer}
\affiliation{Department of Applied Physics and Ginzton Laboratory, Stanford University, 348 Via Pueblo Mall, Stanford, California 94305, USA}
\author{Amir H. Safavi-Naeini}%
\email{safavi@stanford.edu}
\affiliation{Department of Applied Physics and Ginzton Laboratory, Stanford University, 348 Via Pueblo Mall, Stanford, California 94305, USA}

\date{\today}%

\begin{abstract}
Efficient interconversion of both classical and quantum information between  microwave and optical frequency is an important engineering challenge. The optomechanical approach with gigahertz-frequency mechanical devices has the potential to be extremely efficient due to the large optomechanical response of common materials, and the ability to localize mechanical energy into a micron-scale volume. However, existing demonstrations suffer from some combination of low optical quality factor, low electrical-to-mechanical transduction efficiency, and low optomechanical interaction rate. Here we demonstrate an on-chip piezo-optomechanical transducer that systematically addresses all these challenges to achieve nearly three orders of magnitude improvement in conversion efficiency over previous work. Our modulator demonstrates acousto-optic modulation with $V_{\pi} = \SI{0.02}{\volt}$. We show bidirectional conversion  efficiency of $10^{-5}$ with $\SI{3.3}{\micro\watt}$ red-detuned optical pump, and $ 5.5\% $ with $\SI{323}{\micro \watt}$ blue-detuned pump. Further study of quantum transduction at millikelvin temperatures is required to understand how the efficiency and added noise are affected by reduced mechanical dissipation, thermal conductivity, and thermal capacity.
\end{abstract}

\maketitle

It takes energy to sufficiently change the optical properties of a device or medium to impart information onto an optical field~\cite{saleh2019fundamentals,capmany2007microwave}. Electro-optic devices are engineered to minimize this energy by using low-loss optical waveguides and resonators that localize the optical field in a small volume and reduce the amount of energy it takes to setup the electric fields needed for modulation~\cite{miller2017attojoule,marpaung2019integrated}. Mechanical vibrations change the local optical properties with less energy than is typically possible via the electro-optic effect in common materials. Whereas voltages corresponding to $\approx 10^{10}$ microwave photons are typically needed in the most highly optimized electro-optic systems, only $\approx 10^4$ microwave phonons of the same energy are needed in the best optomechanical systems~\cite{Safavi-Naeini2019}. However, this efficient modulation requires localization of both optical and mechanical energy into a wavelength-scale volume~\cite{Safavi-Naeini2019}. This complicates electrical driving of this localized mechanical motion and must be addressed by careful co-engineering of a piezo-optomechanical system.

An energy-efficient modulator, whether electro-optic or piezo-optomechanical, also operates as a quantum transducer between microwaves and light where a large optical pump coherently and reversibly couples an optical sideband to microwave-frequency excitations~\cite{safavi2011proposal,tsang2010cavity,Aspelmeyer2014}. Such transducers may one day enable quantum networks that perform distributed quantum sensing and information processing~\cite{kimble2008quantum,stannigel2010optomechanical}. 
There are many physical mechanisms that mediate the exchange of microwave and optical photons~\cite{lambert2019coherent}. Among these, as in the classical case, mechanically mediated conversion offers strong coupling rates and low energy consumption~\cite{midolo2018nano,Safavi-Naeini2019,wu2019microwave}. Electro-optomechanical conversion using MHz-frequency mechanical membranes~\cite{bagci2014optical, andrews2014bidirectional} to mediate interactions between a Fabry-P\'erot cavity and a superconducting microwave resonator has been demonstrated with $47\%$ efficiency and 38 added noise photons~\cite{higginbotham2018harnessing}. Desire for larger conversion rates, lower added noise, and lower heating have motivated investigations of integrated gigahertz-frequency devices which require less optical pump power to operate. Several approaches using aluminum nitride (AlN)~\cite{bochmann2013nanomechanical, fan2013aluminum, vainsencher2016bi, fan2019spectrotemporal}, silicon \cite{Kalaee2019,VanLaer2018}, gallium arsenide (GaAs)~\cite{balram2016coherent, Balram2017Acousto, forsch2018microwave}, and lithium niobate (LN)~\cite{Liang2017,Jiang2019Lithium,shao2019microwave} have been pursued, but the best end-to-end conversion efficiencies have remained on the order of $10^{-8}$~\cite{vainsencher2016bi}.

\begin{figure*}[t]
\centering
\includegraphics[scale=0.5]{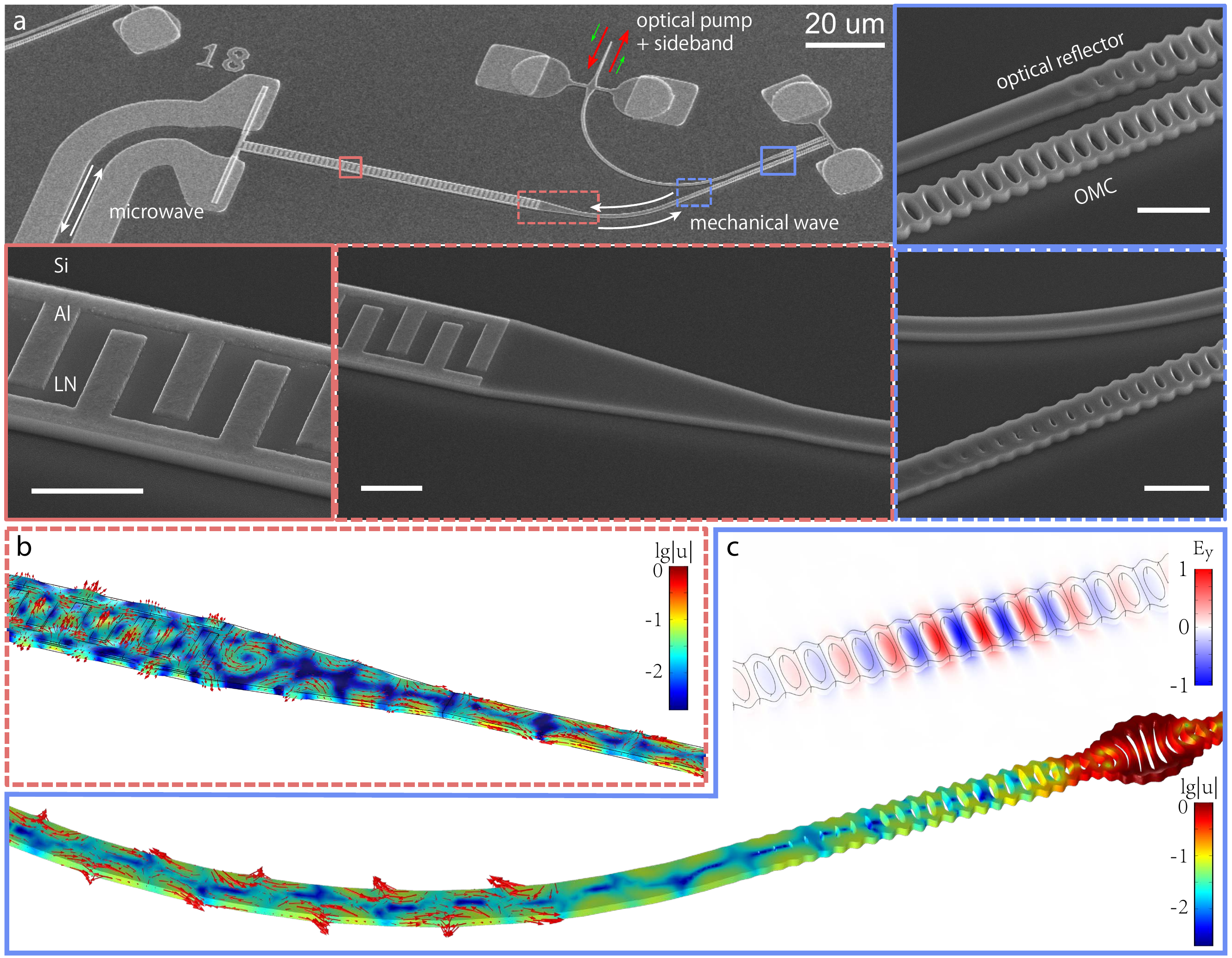} 
\caption{\label{fig1:design} \textbf{Design of lithium niobate piezo-optomechanical transducers.} \textbf{a}, Scanning electron micrographs (SEM) of one piezo-optomechanical transducer. Zoomed-in SEMs show the conversion region between microwave and mechanics (red) and between mechanics and optics (blue). The tapers between the IDT and the mechanical waveguide (dashed red) and between the optomechanical crystal and the waveguide (dashed blue) are also shown in detail. All scale bars in zoomed-in SEMs are \SI{2}{\micro \meter}. \textbf{b}, Finite element simulation of the IDT-to-waveguide taper (normalized displacement). \textbf{c}, Finite element simulations of the OMC optical mode (transverse $E$ component) and the leaky mechanical mode (normalized displacement). First order longitudinal motion in the waveguide can be observed for both the IDT and the leaky OMC mechanical mode.}
\end{figure*}

Efficient piezo-optomechanical modulation is challenging. The mechanical modes must be highly co-localized with the optical resonances to achieve high optomechanical interaction rates, while maintaining electrical access to the mechanical motion. Optomechanical crystals (OMC) provide a natural way to achieve the former by implementing a simultaneous photonic-phononic crystal to confine both optical and mechanical waves~\cite{eichenfield2009optomechanical}. However, efficient electrical coupling to the micron-scale mechanical resonances of a phononic crystal has only recently been achieved~\cite{Arrangoiz-Arriola2018,arrangoiz2019resolving}. These demonstrations leverage the high piezoelectric coefficient of lithium niobate~\cite{Weis1985} and electrodes on or near the resonator to efficiently couple motion to high-impedance superconducting microwave circuits. Such an approach is difficult to integrate with photonic devices due the large optical absorption of metals, which ruins the optical quality factor and destroys the superconductivty.

We overcome the low microwave-to-mechanical efficiency of previously demonstrated lithium niobate piezo-optomechanical crystals~\cite{Jiang2019Lithium} by integrating an interdigitated transducer (IDT) that excites a wavelength-scale mechanical waveguide~\cite{dahmani2019piezoelectric} to efficiently drive the localized mechanical mode of the OMC. Moreover, the phononic waveguide spatially separates the optical and microwave circuits, an important feature in a cryogenic setting where absorption in metals needs to be minimized. We characterize the device as a classical modulator where an incoming microwave signal at the mechanical mode frequency  modulates the optical cavity resonance. In a separate experiment we characterize the potential of the device as a quantum transducer, by imparting a laser tone red-(blue-)detuned by a mechanical frequency from the optical resonance to introduce an interaction between photons resonant with the cavity and the mechanical motion of the device, which is coupled to the external microwave channel. We demonstrate bidirectional conversion between microwave and optical photons with efficiency up to $10^{-5}$ ($5.5\%$) using the red-(blue-)detuned optical pump. The integrated piezo-optomechanical transducer is fabricated with $X$-cut thin-film lithium niobate on silicon (LNOS), a material platform demonstrated to be compatible with superconducting circuits and qubits~\cite{Arrangoiz-Arriola2018,arrangoiz2019resolving} -- opening a path for integration with quantum sensors and processors.

\section*{Results}

\textbf{Design.} An incident microwave signal on the IDT is converted to a propagating mechanical wave in the second-order horizontal shear mode (SH2) in the transducer region.  The mechanical wave is then scattered by the linear horn into the first-order longitudinal mode (L1) of a \(\SI{1.3}{\micro\meter}\)-wide waveguide (Fig.~\ref{fig1:design}b). From finite-element simulation and separate measurement~\cite{dahmani2019piezoelectric}, we determine that $\sim 0.8\%$ of the microwave power absorbed by the IDT is converted to mechanical motion in the waveguide, the rest of which is lost to dissipation and clamping. From this fraction, $75\%$ is in the L1 mode leading to a $0.6\%$ conversion from microwave input to mechanical power in the L1 mode for a perfectly matched IDT. Phonons propagate down the waveguide and are scattered into the guided mode of the OMC by smoothly ramping on the OMC's periodic modulation.
Scattering of mechanical waves from the waveguide into the localized mechanical mode can be induced by breaking a symmetry of the structure. This occurs automatically in our device, in contrast to previous work on silicon OMCs~\cite{patel2017engineering}, which required patterns that explicitly broke symmetry to induce scattering. This is because the lithium niobate material lacks crystal symmetry along the reflection plane of the OMC geometry~\cite{Jiang2019Lithium}. We simulate a decay rate $\gammae/2\pi \approx \SI{460}{\kilo\hertz}$ using COMSOL~\cite{COMSOL}, $64\%$ of which is converted to the L1 mode (Figure~\ref{fig1:design}(c)). The IDT pitch \(a \approx \SI{2.68}{\micro\meter}\) is chosen to match the frequencies of the IDT and OMC. Since LN is anisotropic, both the IDT and the OMC have optimal orientations where the piezoelectric and photoelastic effects are respectively maximized. A curved waveguide with a bending radius of $\SI{30}{\micro\meter}$ connects the two components. 

Optical photons are injected into an on-chip edge coupler via a lensed fiber with coupling efficiency $\eta_{\textrm{oc}}\sim 65\%$. This waveguide is brought in the near-field of the OMC to allow coupling of light fields in and out of the optical resonance. The fabrication process is described in Ref.~\cite{Jiang2019Lithium,dahmani2019piezoelectric}. The scanning electron micrographs (SEM) in Fig.~\ref{fig1:design} were taken before the final masked release step.

\begin{figure*}[tb]
\centering
\includegraphics[scale=0.33]{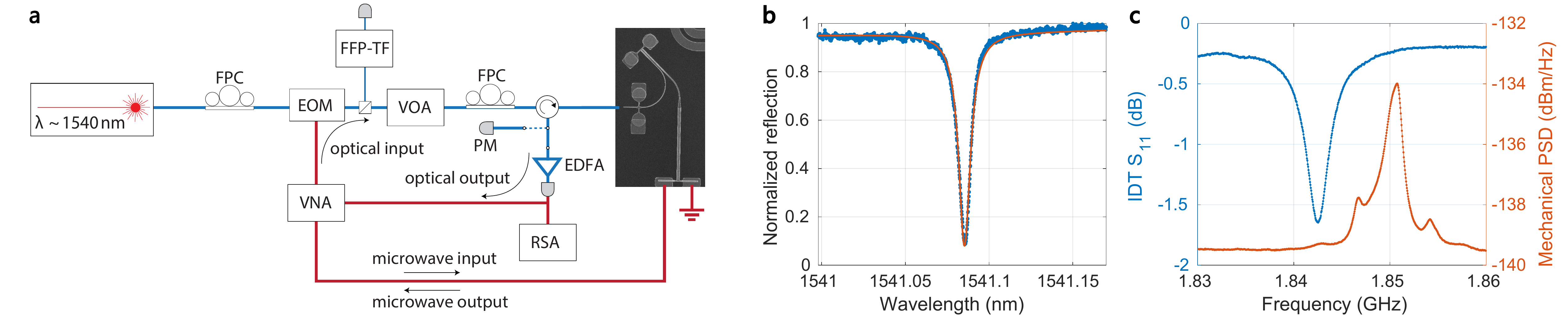}
\caption{\label{fig2:setup-basic-meas} \textbf{Measurement setup and separate characterizations of the IDT and the OMC.} \textbf{a}, Schematic drawing of the measurement setup. The optical sideband input is generated by a microwave tone from the vector network analyzer (VNA) to the electro-optic modulator (EOM). Optical readout of the sideband is generated by a highspeed photodetector and is sent back to the VNA. Microwave input and output of the device is directly actuated by and measured from the IDT. \textbf{b}, Optical resonance of the optomechanical crystal. \textbf{c}, IDT response (blue) and OMC mechanical mode thermal spectroscopy (red). Variable optical attenuator (VOA), fiber polarization controller (FPC), erbium-doped fiber amplifier (EDFA), realtime spectrum analyzer (RSA), power meter (PM), fiber Fabry-P\'erot tunable filter (FFP-TF).}
\end{figure*}

\begin{figure*}[tb]
\centering
\includegraphics[scale=0.4]{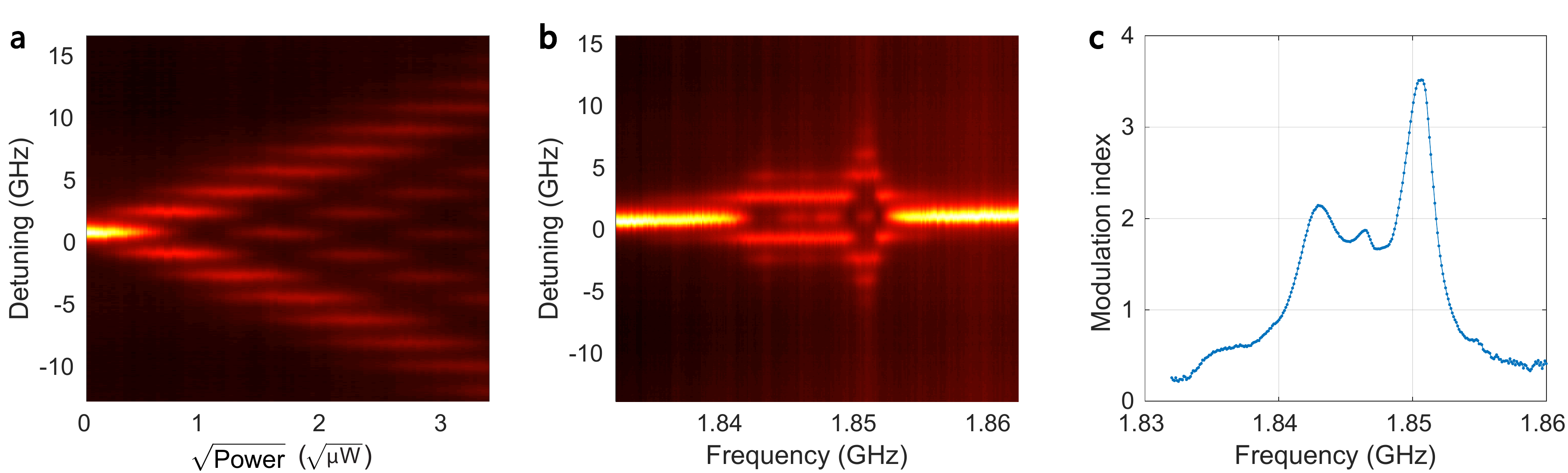}
\caption{\label{fig3:AOM} \textbf{Efficient acousto-optic modulation.} \textbf{a, b,} A microwave signal sent to the IDT modulates the optical cavity frequency. The reflection spectrum of the optical cavity is recorded for different microwave power (\textbf{a}) and frequency (\textbf{b}). \textbf{c}, Modulation index for different microwave drive frequency, extracted from (b).}
\end{figure*}

\textbf{Device characterization.} Figure~\ref{fig2:setup-basic-meas}(a) shows the measurement setup used for this work. We use a vector network analyzer (VNA) to generate and read out signals to and from the IDT (microwave input and output). A commercial electro-optic modulator (EOM) is driven by the VNA to generate the optical sideband input (optical input). The light reflected from the device containing the pump and sideband is amplified and collected by a highspeed detector, which downconverts the optical sideband to a microwave signal received by the VNA (optical output). We measure all four scattering parameters of this two-port setup at room temperature. All ports are calibrated to de-embed the scattering parameters of the device.

We begin by characterizing the OMC and the IDT. A Lorentzian response is observed when we scan a tunable telecom laser with frequency $\omegap$ across the optical resonance of the OMC at frequency $\omegac$ (Fig.~\ref{fig2:setup-basic-meas}(b)). Combined with the optical sideband response, we extract a total optical linewidth $\kappa/2\pi = 1210\pm 40~\textrm{MHz}$ and an external coupling rate $\kappae/2\pi = 800 \pm 30~\textrm{MHz}$, corresponding to a loaded quality factor $Q = 1.6\times 10^5$ and an intrinsic quality factor $Q_{\textrm{i}} = 4.7\times 10^5$. The IDT response is measured with a calibrated microwave probe~\cite{dahmani2019piezoelectric} and is shown in Fig.~\ref{fig2:setup-basic-meas}(c). We achieve a peak conductance of $1.9~\textrm{mS}$ and a bandwidth of $ B_{\textrm{IDT}} = 3.36~\textrm{MHz} $ with direct coupling to a $\SI{50}{\ohm}$ transmission line. Lastly we characterize the mechanical mode and optomechanical coupling by thermal spectroscopy and optomechanical backaction~\cite{Aspelmeyer2014,Jiang2019Lithium}. A typical thermal mechanical power spectral density (PSD) is plotted in Fig.~\ref{fig2:setup-basic-meas}(c) in red. The fundamental mechanical breathing mode is at $ \omegam/2\pi = 1.85~\textrm{GHz}$. Satellite peaks are visible and indicate the coupling between the local breathing mode and mechanical waveguide modes. We measure a backaction-free mechanical linewidth $\gamma/2\pi = 1.93~\textrm{MHz}$ and an optomechanical coupling rate $g_{0}/2\pi = 70~\textrm{kHz}$ via the optomechanical backaction (Supplementary Information). We note that both $\gamma$ and $g_{0}$ differ significantly from the values measured in Ref.~\cite{Jiang2019Lithium} with identical OMC geometry and orientation. We attribute these deviations to hybridization of the local breathing mode with the waveguide modes.

\textbf{Efficient acousto-optic modulation.} To demonstrate acousto-optic modulation, we drive the mechanical mode of the OMC with a microwave tone at frequency $\omegamu \sim \omegam$ and measure the reflection spectrum of the optical mode. The coherent mechanical motion induces a varying optical cavity frequency 
\begin{equation}
    \widetilde{\omegac} = \omegac +  \widetilde{\Delta \omegac} =\omegac+ g_{0} \sqrt{\nphon} \cos \omegamu t,
\end{equation}
where $\nphon$ is the driven intracavity phonon number proportional to the input microwave power $\Pmu $. This phase modulation splits the optical cavity spectrum into sidebands at $\omegac \pm n \omegamu$ with integer sideband index $n$ and relative strength determined by $n$ and the modulation index $h \equiv g_{0} \sqrt{\nphon} /\omegamu $ (Supplementary Information). We measure the cavity reflection spectrum for different microwave powers and frequencies. 

Figure~\ref{fig3:AOM}(a) shows the reflection spectrum versus cavity-laser detuning $\Delta = \omegac - \omegap$ and microwave power $\Pmu$. The phase modulation sidebands are clearly resolved. We empirically confirm the expected proportionality between the square root of the microwave power and the modulation index $h$. We show the measured spectrum with fixed power $\Pmu = \SI{7.24}{\micro\watt}$ at different frequencies in Fig.~\ref{fig3:AOM}(b) and extract the modulation index versus frequency in Fig.~\ref{fig3:AOM}(c) by fitting the spectrum (Supplementary Information). The modulation index peaks at the mechanical mode frequency $\omegamu = \omegam$, where the microwave power required to achieve modulation index $h=1$ is as low as $\Pmu = \SI{0.58}{\micro\watt}$. The complex shape of the excitation spectrum in Fig.~\ref{fig3:AOM}(c) is due to the mismatch of the IDT and mechanical mode frequencies (see Fig.~\ref{fig2:setup-basic-meas}(c)). The lower local maxima match the IDT response and standing wave modes of the mechanical waveguide. The mismatch between the local mechanical mode frequency and the IDT frequency lowers the peak modulation, but increases the $3~\textrm{dB}$ bandwidth to $B \sim 10~\textrm{MHz}$. Combining the acousto-optic modulation measurement with $g_0$ and $\gamma$, we deduce a decay rate from the mechanical mode to the $Z_0 = \SI{50}{\ohm}$ microwave transmission line of $\gammamu/2\pi = 8.6~\textrm{kHz}$ and the corresponding microwave-to-mechanical conversion efficiency $ \etam \equiv \gammamu/\gamma = 0.44\%$ (Supplementary Information).

An important figure of merit for a classical modulator is the amount of driving energy required to encode one bit of information on an optical field~\cite{miller2017attojoule}. This generally requires consideration of multiple aspects of a communication system. We consider a simplified thought experiment, where we begin with the optical resonator prepared initially in a coherent state $|\alpha_0|^2=1$ and ask how well can the output optical field $|{\Psi_k}\rangle$ be distinguished when the modulator is driven to encode $k=0$ or $k=1$. In the supplementary information we use the Helstrom-Holevo bound~\cite{wiseman2009quantum} to calculate the driving energy required to realize a bit-error-rate of $10\%$ for different system parameters $(g_0,\etam,\kappa,\omegam)$ and find two limiting cases:
\begin{eqnarray}
    E_\text{bit} &=& \frac{\hbar\omegam}{4\etam}\left(\frac{\kappa}{g_0}\right)^2~~\text{for}~\kappa\gg\omegam,~~
    \text{and}\label{eqn:Ebitslow}\\
    E_\text{bit} &=& \frac{\hbar\omegam}{2\etam}\left(\frac{\omegam}{g_0}\right)^2~~\text{for}~\kappa\ll\omegam. \label{eqn:Ebitfast}
\end{eqnarray}
In our sideband-resolved system ($\kappa<\omegam$), the second equation leads to an energy-per-bit of $\SI{97}{\femto\joule}$. Notice that a modulation index of $h=\pi$ is achieved with $V_\pi=\SI{24}{\milli\volt}$, requiring and RF power $P_\pi=V_\pi^2/2Z_0$ which can also be used to estimate the energy-per-bit $P_\pi/(2\pi B)\sim \SI{100}{\femto \joule}$.

\begin{figure*}[tb]
\centering
\includegraphics[scale=0.4]{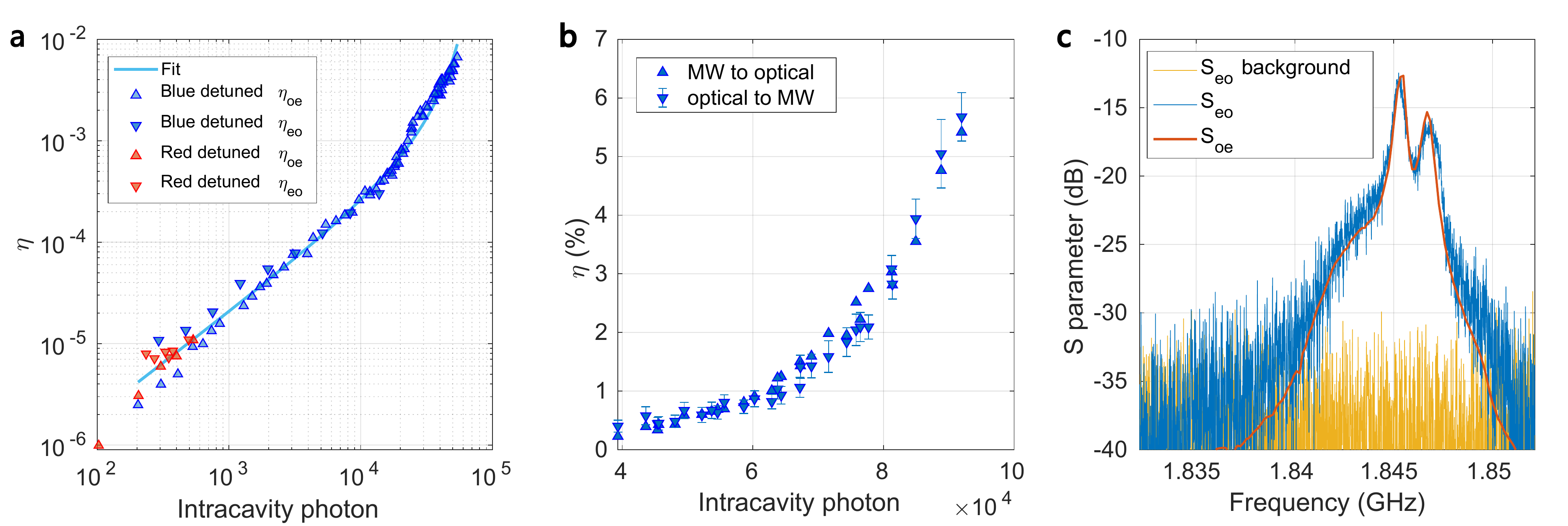}
\caption{\label{fig4:EO-conversion} \textbf{Bidirectional microwave-optical conversion.} \textbf{a}, Conversion efficiency measured with red and blue detuned pump laser. Upward (downward) triangles represent the microwave-to-optical (optical-to-microwave) conversion efficiencies, where the frequency of the photon is up-converted (down-converted). The optical-to-microwave efficiency $\eta_{\textrm{eo}}$ is measured with a different setup. \textbf{b}, Conversion efficiency measured with relatively high power blue-detuned pump laser using same measurement setup for $ \eta_{\textrm{oe}} $ and $ \eta_{\textrm{eo}} $ (Fig.~\ref{fig2:setup-basic-meas}(a)). \textbf{c}, Bidirectional scattering parameter measurement. The maximal $|S|^2$ is shifted to the calibrated efficiency. The $S_{\textrm{eo}}$ background is measured by sending the VNA output to a $\SI{50}{\ohm}$ load instead of the EOM.}
\end{figure*}

\textbf{Bidirectional conversion between microwave and optical frequency.} We demonstrate microwave-to-optical conversion by pumping the optical mode with detuning $ \Delta = \pm \omegam $ using different pump powers and measuring the microwave-to-optical scattering parameter. The total conversion efficiency is defined as the ratio between input microwave photon flux before the IDT and output sideband photon flux after the light being collected by the lensed fiber. At $\Delta = \pm \omegam$, the efficiency is given by
\begin{equation}
\label{eqn:etaOE}
    \eta_{\textrm{total}} = \eta_{\textrm{oc}} \etao \etam \frac{4C}{(1\pm C)^2},
\end{equation}
where $\eta_{\textrm{oc}} = (65.2\pm0.4) \%$ is the fiber-to-chip optical coupling efficiency and $\etao \equiv \kappae/\kappa=(66\pm 1)\% $ is the external optical coupling efficiency. $C = C_{0} \nc$ is the optomechanical cooperativity, $\nc$ is the intracavity photon number and $C_{0} \equiv 4g_0^2/(\kappa \gamma)$ is the single-photon cooperativity.

For a red-detuned pump, the linearized optomechanical interaction gives a beam-splitter  interaction Hamiltonian for a sideband-resolved system~\cite{Aspelmeyer2014}. Ideal internal quantum transduction can be achieved in principle at the matching condition $C = 1$, where the maximal efficiency equals the total external coupling efficiency $\eta_{\textrm{total}} = \etae \equiv \eta_{\textrm{oc}} \etao \etam  $. Alternatively, a blue-detuned optical pump leads to a two-mode squeezing Hamiltonian that could be utilized to generate entanglement between optical and microwave photons, required for long range quantum communication~\cite{duan2001long,zhong2019heralded}. It also gives rise to phase-conjugating amplification of input optical sideband photons or input microwave photons with added noise~\cite{caves1982quantum}. The internal gain $G_{\textrm{int}} \equiv 4C/(1-C)^2$ monotonically increases with optomechanical cooperativity $C$ until the phonon lasing condition $C \ge 1$, where $G_{\textrm{int}} \rightarrow \infty $ and the system no longer operates in the linear regime. In our demonstration, we reached  $G_{\textrm{int}}>1 $ compensating for some, but not all of the external coupling losses $\etae$. We refer to the scattering parameter measurement under blue-detuned pump as ``efficiency'' instead of ``gain'' with the understanding that this is not a quantum-state-conversion process.

We calibrate the optical detection gain using a second calibration laser (see Supplementary Information and also Ref.~\cite{patel2018single} for detailed description). The calibrated total microwave-to-optical conversion efficiency $\eta_{\textrm{oe}}$ is shown in Fig.~\ref{fig4:EO-conversion}(a) for different intracavity photon numbers. Measurements with a red-detuned pump are limited  by the thermo-optical instability~\cite{Jiang2019Lithium} to intracavity photon $\nc \lesssim 500$ and a resulting total efficiency $\eta_{\textrm{oe}} \sim 10^{-5}$. We measure the efficiency with a blue-detuned pump up to $\nc \sim 4\times 10^4$ and fit to Eq.~\ref{eqn:etaOE} with two fitting parameters $\etae  $ and $C_{0}$. The fit curve is plotted in Fig.~\ref{fig4:EO-conversion}(a) in cyan, showing good agreement with measured efficiencies across more than three orders of magnitude. We extract a single-photon cooperativity $C_{0} = 1.2\times 10^{-5}$ and a total external coupling efficiency $\etae = 4.24\times 10^{-4}$. The deviation of measured efficiencies from the fit at high $\nc$ can be attributed to thermally-induced redshifts of the mechanical mode (Supplementary Information). We also calculate $ C_{0} = 8.4\times 10^{-6}$ from independent measurement of $g_0 $, $\kappa$ and $\gamma $. The $g_{0}$ obtained from blue-detuned backaction measurement is possibly underestimating the actual $g_{0}$ due to thermal broadening of the mechanical linewidth~\cite{Jiang2019Lithium}. We deduce $g_0/2\pi = 84~\textrm{kHz}$ with $C_{0}$ value obtained from the efficiency measurement. After removing the independently measured $\etao $ and $\eta_{\textrm{oc}}$ from $\etae$, we calculate $\etam = 9.9\times 10^{-4} \sim 0.1\%$ and the corresponding $\gammamu/2\pi = \SI{1.9}{\kilo\hertz}$, about five orders of magnitude increase from our previous demonstration~\cite{Jiang2019Lithium}. Improving the IDT-OMC mechanical frequency mismatch of $\Delta f = 8.3~\textrm{MHz} \sim 2.5 B$, would increase $\etam$ by an order of magnitude. We note a factor of $ 4.4 $ difference between $\etam$ measured from acousto-optic modulation and $\etam$ measured from conversion efficiencies. An overestimated $g_{0}$, as well as device aging between measurements affecting the frequency mismatch, explains this discrepancy in $\gammamu$ (Supplementary Information).

In optical-to-microwave conversion, due to the much lower energy per photon at microwave frequency, a $\sim -50~\textrm{dB}$ reduction of the microwave signal power with respect to optical is incurred. With sub-microwatt optical sideband power, it becomes technically challenging to keep the converted microwave power above the VNA internal crosstalk level. We adopt a more  sensitive measurement scheme~\cite{VanLaer2018} that uses a signal generator for the EOM input and an isolated realtime spectrum analyzer (RSA) for sensitive detection of the microwave output. The optical input sideband is calibrated by a fiber Fabry-P\'erot tunable filter (FFP-TF) and is described in detail in Supplementary Information. The calibrated optical-to-microwave conversion efficiencies $\eta_{\textrm{eo}}$ are plotted in Fig.~\ref{fig4:EO-conversion}(a) for comparison. Higher efficiencies are observed for blue-detuned $\eta_{\textrm{eo}}$ than $\eta_{\textrm{oe}}$. We attribute the difference to temperature fluctuation and device aging during the modification of measurement setup, which could lead to a different frequency mismatch between the OMC mechanical mode and a nearby waveguide mode.

\begin{table*}
\caption{\textbf{Comparison of integrated piezo-optomechanical transducers.} $\gammamu $ is defined as the decay rate from the local mechanical mode to the $\SI{50}{\ohm} $ microwave channel. Optical-to-microwave ($\eta_{\textrm{eo}}$), microwave-to-optical ($\eta_{\textrm{oe}}$) and blue-side-pump ($\eta_{\textrm{blue}}$) efficiencies are listed separately. We note that only the demonstrations on LN platform are in the sideband-resolved regime, and the corresponding $E_{\textrm{bit}}$ is calculated differently (see the main text). The best quantities are highlighted in bold.} \label{tab:parameters}
\begin{center}
\begin{tabular}{c| p{2.2cm} p{2cm} p{2.3cm} p{2.2cm} p{2.2cm}| p{2.3cm} }
Reference: & Vainsencher \textit{et al.} \cite{vainsencher2016bi} & Balram \textit{et al.} \cite{balram2016coherent,Balram2017Acousto}& Forsch \textit{et al.}~\cite{forsch2018microwave} & Jiang  \textit{et al.}~\cite{Jiang2019Lithium} & Shao  \textit{et al.}~\cite{shao2019microwave} & This work \\ 
\hline
\hline
Platform: & AlN  & GaAs & GaAs ($20$ mK) & LN & LN & LN\\
$g_0/2\pi$ (Hz) & $1.1\times 10^5$ & $1.1\times 10^6$ &  $ \bm{ 1.3\times 10^{6} } $  & $1.2\times 10^5$ & $1.1\times 10^3$ & $8\times 10^4$ \\
$ \kappa/2\pi $ (Hz) & $1.5\times 10^{10} $ & $5.2\times 10^9$ & $5.8\times 10^9 $ & $7.8\times 10^8$ & $ \bm{ 9.5\times 10^7 }$ & $1.2\times 10^9$ \\
$ \gamma/2\pi $ (Hz) & $5\times 10^{6} $ & $1.7\times 10^6$ & $\bm{ 2\times 10^5 } $  & $5\times 10^5$ & $1.3\times 10^6$ & $1.9\times 10^6$ \\
$ C_{0}\equiv 4g_0^2/(\kappa\gamma) $ & $7\times 10^{-7} $ & $5.4\times 10^{-4}$ & $ \bm{ 5.9\times 10^{-3} } $ & $1.5\times 10^{-4}$ & $4\times 10^{-8}$ & $1.2\times 10^{-5}$ \\
$\etao \equiv \kappae/\kappa$  & $\bm{ 69\% }$ & $ 23\% $ & $ 65\% $ &  $29\%$  & $15\%$ &  $66\%$ \\
$ \etam \equiv \gammamu/\gamma $ & $\sim 2\times 10^{-4} $ & $3\times 10^{-10}$ & $ 3.6\times 10^{-10} $ & $1.8\times 10^{-8}$ & $\bm{1.7\times 10^{-1}} $ & $10^{-3}$ \\
$\eta_0 \equiv 4 \etao \etam C_{0} $ & $ \sim 3.7\times 10^{-10} $ & $ 1.5\times 10^{-13}$ & $ 5.5\times 10^{-12} $ & $3\times 10^{-12}$ & $ 4\times 10^{-9}$ & $ \bm{3.2 \times 10^{-8} }$ \\
\hline
$ \eta_{\textrm{int}}\equiv 4C/(1+C)^2 $ & $ 6.5\times 10^{-3}  $ & -- & $\bm{7.2\times 10^{-2}} $ & $\sim 10^{-2} $ & $7\times 10^{-4} $ & $ 2.6\times 10^{-2} $ \\
$ \eta_{\textrm{eo}} $ & $ 9\times 10^{-8}  $ & -- & -- & -- & -- & $ \bm{1.1\times 10^{-5}} $ \\
$ \eta_{\textrm{oe}} $ & $2\times 10^{-8}$ & -- & $ 5.5\times 10^{-12} $ & {$\sim 10^{-10}$}  & -- &  {$\bm{1.1\times 10^{-5} } $} \\
$ \eta_{\textrm{blue}} $ & -- & -- & -- & -- & {$1.7\times 10^{-5}$ } &  { $\bm{5.5\times 10^{-2}}$}\\
$ P_{\textrm{pump}} $  & $\sim \SI{60}{\micro\watt}
~(\eta_{  \textrm{eo}}) $  $\sim \SI{110}{\micro\watt}~(\eta_{ \textrm{oe}}) $ & -- & ${\bm{\sim 0.5}~\SI{}{\micro\watt}} $ &{$\sim \SI{3}{\micro \watt}$} & {\SI{1}{\milli\watt}} & {\SI{3.3}{\micro \watt}}~(red)  {\SI{323}{\micro \watt} }~(blue)\\
$ E_{\textrm{bit}} $ (\SI{}{\joule}) & $5.7\times10^{-11} $ & $3\times10^{-8} $ &$2.5\times10^{-8} $ & $8.1\times10^{-9} $ & $5.6\times10^{-11} $ & $\bm{9.7\times10^{-14}} $ \\
$ E_{\textrm{qubit}} $ (\SI{}{\joule}) & $1.3\times10^{-6} $ & $8.3\times10^{-3} $ & $9.7\times10^{-4} $ & $1.9\times10^{-4} $ & $7.8\times10^{-9} $ & $\bm{3.5\times10^{-9} }$ \\
\hline
\hline
\end{tabular}
\end{center}
\end{table*}

To achieve bidirectional conversion measurement with the VNA, we fix the maximal pump laser power to $ \sim \SI{300}{\micro\watt}$ limited by the setup and change the detuning $2\pi\times 1~\textrm{GHz} < |\Delta| \leq \omegam $ to further increase the intracavity photon number $\nc$. The measured conversion efficiencies are shown in Fig.~\ref{fig4:EO-conversion}(b). Reasonable agreement between the two conversion directions is observed up to the phonon lasing regime, with highest stable conversion efficiency $\eta \approx 5.5\%$ in the linear operation regime. The corresponding internal gain is $G_{\textrm{int}} \approx 21~\text{dB}$. Figure~\ref{fig4:EO-conversion}(c) shows the measured $S$ parameters at the highest achieved efficiency. We obtain good agreement between optical-to-microwave and microwave-to-optical scattering parameters for both the lineshape and the peak efficiency. When $\nc$ is further increased such that $C>1$, we observe phonon lasing and amplification for both input optical sideband and input microwave signal. Operation in this nonlinear regime will be studied in future work.

\section*{Discussion}

To compare existing piezo-optomechanical transducers, we summarize parameters from various demonstrations in Table~\ref{tab:parameters}. $\etam$ and $P_{\textrm{pump}}$ from Vainsencher \textit{et al.}~\cite{vainsencher2016bi} are estimated from their measured total efficiency, intracavity photon number and other system parameters, assuming a grating coupler efficiency of $\sim 20\%$. The $\etam$ from Balram \textit{et al.}~\cite{balram2016coherent} is calculated from the measured driven coherent phonon population. We note that Shao \textit{et al.}~\cite{shao2019microwave} achieved an efficiency of $17\%$ between microwave and mechanics, orders of magnitude higher than other approaches but at the cost of a reduced overlap between the optical and mechanical modes leading to a lower optomechanical coupling rate $g_{0}/2\pi=\SI{1.1}{\kilo\hertz}$. They estimated a conversion efficiency $\eta_{\textrm{blue}} = 1.7\times 10^{-5} $ from their system parameters. We define the on-chip single-pump-photon conversion efficiency as $\eta_0 \equiv 4 \etao\etam C_0$. The transducer reported here achieved roughly one order of magnitude higher $\eta_0$ comparing to all other demonstrations. The fiber-to-chip optical coupling efficiency $\eta_{\textrm{oc}}$ is a separate issue and is not explicitly included in Table~\ref{tab:parameters}, though it affects the total efficiency numbers.

As a modulator for classical optical fields, our device exhibits both a $V_\pi$ and an energy-per-bit $E_\text{bit}$~\cite{miller2017attojoule,Safavi-Naeini2019} that is orders of magnitude lower than previous acousto-optic demonstrations. We use equations (\ref{eqn:Ebitslow}) and~(\ref{eqn:Ebitfast}) to estimate the energy-efficiency of a range of devices. In general, the piezo-optomechanical devices lag significantly behind the best electro-optic devices \cite{Reed2010,wang2018integrated,Kieninger2018,Abel2019}. By further improving the electrical-to-mechanical efficiency from $\etam\sim 10^{-3}$ to closer to unity, we expect to push the classical performance of the device deep into the sub-femtojoule regime where performance in excess of the electro-optic systems becomes possible. Implementing a similar approach in a hybrid platform that integrates silicon with lithium niobate~\cite{witmer2017high} would allow even greater improvements by increasing $g_0$ by an order of magnitude and enabling sub-attojoule modulation energy. Finally, we stress that the low dissipated energies-per-bit are not directly related to the limited mechanical bandwidth but rather mostly as a result of the strong optomechanical interaction. The driving bandwidth could be increased significantly by advances in OMC and IDT design for applications requiring faster modulation while keeping a low dissipated energy-per-bit.

For quantum transduction, each converted qubit comes at a cost of optical pump power dissipated in the fridge $E_{\textrm{qubit}} = \hbar \omegac \kappa \kappai /g_0^2/(4\etao \etam)$, where $\kappai$ is the intrinsic optical cavity loss rate~(see Ref.~\cite{pechal2017millimeter,Safavi-Naeini2019} and also the Supplementary Information). Cooling our transducer to cryogenic temperature would reduce material loss for the IDT and the OMC, and may increase both $\etam $ and $C_{0} $ by more than one order of magnitude. Adding another order of magnitude from matching IDT and OMC mechanical frequencies, we expect a picojoule energy-per-qubit and more than three orders of magnitude better efficiency to be possible, bringing the total efficiency with a red detuned pump up to $\gtrsim 1\%$ with only $\sim 500$ intracavity photons and corresponding optical pump power $ P_{\textrm{pump}} \approx \SI{3.3}{\micro\watt} $. When the transducer is further resonantly coupled to a superconducting qubit/resonator with a characteristic impedance $Z_{\textrm{c}}\sim \SI{300}{\ohm}$, we estimate the coupling rate between the qubit/resonator and the OMC mechanical mode to be $g_{\upmu} = \sqrt{\gammamu \omegam}\cdot \sqrt{Z_{\textrm{c} }/Z_0 }/2 \sim 2\pi \times \SI{2.3}{\mega\hertz}$~\cite{Jiang2019Lithium}, putting us in the strong coupling regime so long as the qubit/resonator linewidth $ \kappa_{\upmu} /2\pi < \SI{2.3}{\mega\hertz} $.

In conclusion, we designed and fabricated an integrated piezo-optomechanical transducer by combining an efficient wavelength-scale mechanical waveguide transducer and an optimized optomechanical crystal on LNOS platform. The microwave-to-mechanical conversion efficiency is increased by a factor of $10^5 $ comparing to our previous design without severely impacting the optomechanical coupling or dissipation. We demonstrated efficient acousto-optic modulation with $V_{\pi} = 24~\textrm{mV}$, bidirectional conversion efficiency of $10^{-5}$ with $\SI{3.3}{\micro\watt}$ red-detuned optical pump and $ 5.5\% $ with $\SI{323}{\micro \watt}$ blue-detuned pump at room temperature.  We expect our transducers to have reduced material loss and an increased efficiency at cryogenic temperature, opening up experiments in the quantum regime between optical photons, microwave photons and phonons, and superconducting qubits~\cite{Arrangoiz-Arriola2018, arrangoiz2019resolving}.

\bibliography{LINQS_LNPOMT}

\section*{Acknowledgment}

 W.J. would like to thank Jeremy D. Witmer, Patricio Arrangoiz-Arriola and Agnetta Y. Cleland for helpful discussions. This work is supported by National Science Foundation (NSF) (1708734, 1808100), Army Research Office (ARO/LPS) (CQTS), Airforce Office of Scientific Research (AFOSR) (MURI No. FA9550-17-1-0002 led by CUNY), Fonds Wetenschappelijk Onderzoek (FWO Marie Sklodowska-Curie grant agreement No. 665501), VOCATIO. R.N.P. is partly supported by the NSF Graduate Research Fellowships Program. Device fabrication was performed at the Stanford Nano Shared Facilities (SNSF) and the Stanford Nanofabrication Facility (SNF). SNSF is supported by the National Science Foundation under award ECCS-1542152.

\section*{Author contributions}
W.J., Y.D.D. and C.J.S. designed the device. W.J. fabricated the device assisted by F.M.M.. W.J., T.P.M., F.M.M. and R.V.L. developed the fabrication process. W.J. measured the device. R.N.P., F.M.M. and C.J.S. provided assistance with the measurements. C.J.S. and Y.D.D. performed the mechanical mode decomposition simulation with assistance from W.J.. W.J. wrote the manuscript with input from all authors. A.H.S.-N. supervised the project.

\section*{Competing interests}
The authors declare no competing interests.

\onecolumngrid

\newcommand{\omegak}{\omega_{k}}
\newcommand{\omegaL}{\omega_{\textrm{L}}}
\newcommand{\omegaFSR}{\omega_{\textrm{FSR}}}
\newcommand{\opdagger}[2]{\mbox{$\hat{#1}_{#2}^{\dagger}$}}
\newcommand{\opd}[2]{\mbox{$\hat{#1}_{#2}^{\dagger}$}}  %
\newcommand{\op}[2]{\mbox{$\hat{#1}_{#2}$}}

\newcommand{\kappaemu}{\kappa_{\textrm{e,}\mu}}
\newcommand{\kappamu}{\kappa_{\mu}}
\newcommand{\kappamue}{\kappa_{\mu,\textrm{e}}}
\newcommand{\gmu}{g_{\mu}}

\newcommand{\ncsty}{\bar{n}_{\textrm{c}}}

\newcommand{\opA}{\hat{A}}
\newcommand{\opAd}{\hat{A}^\dagger}
\newcommand{\cout}{\hat{c}_{\textrm{out}}}
\newcommand{\cin}{\hat{c}_{\textrm{in}}}

\newcommand{\alpham}{\alpha_{-}}
\newcommand{\alphainp}{\alpha_{\textrm{in},+}}
\newcommand{\alphaoutp}{\alpha_{\textrm{out},+}}
\newcommand{\alphainm}{\alpha_{\textrm{in},-}}
\newcommand{\alphain}{\alpha_{\textrm{in}}}
\newcommand{\alphaout}{\alpha_{\textrm{out}}}
\newcommand{\alphainpm}{\alpha_{\textrm{in},\pm}}
\newcommand{\alphapm}{\alpha_{\pm}}
\newcommand{\betap}{\beta_{+}}
\newcommand{\betam}{\beta_{-}}
\newcommand{\betapm}{\beta_{\pm}}
\newcommand{\betainp}{\beta_{\textrm{in},+}}
\newcommand{\betainm}{\beta_{\textrm{in},-}}
\newcommand{\betaoutm}{\beta_{\textrm{out},-}}
\newcommand{\betaoutp}{\beta_{\textrm{out},+}}

\newcommand{\Aa}{A_{\textrm{a}}}
\newcommand{\Ab}{A_{\textrm{b}}}
\newcommand{\Ac}{A_{\textrm{c}}}

\newcommand{\etaab}{\eta_{\textrm{ab}}}
\newcommand{\etaba}{\eta_{\textrm{ba}}}
\newcommand{\etacb}{\eta_{\textrm{cb}}}
\newcommand{\etabc}{\eta_{\textrm{bc}}}
\newcommand{\etabcb}{\eta_{\textrm{bcb}}}
\newcommand{\etaaba}{\eta_{\textrm{aba}}}
\newcommand{\etaabc}{\eta_{\textrm{abc}}}
\newcommand{\etacba}{\eta_{\textrm{cba}}}
\newcommand{\Pinmu}{P_{\textrm{in},\upmu}}
\newcommand{\Gammatot}{\Gamma_{\textrm{tot}}}

\newcommand{\Soe}{S_{\textrm{oe}}}
\newcommand{\Seo}{S_{\textrm{eo}}}
\newcommand{\Soo}{S_{\textrm{oo}}}
\newcommand{\etaoe}{\eta_{\textrm{oe}}}
\newcommand{\etaeo}{\eta_{\textrm{eo}}}

\newcommand{\Ndotino}{\dot{N}_{\textrm{in,o}}}
\newcommand{\Ndotoutmu}{\dot{N}_{\textrm{out},\upmu}}
\newcommand{\Ndotinmu}{\dot{N}_{\textrm{in},\upmu}}
\newcommand{\Ndotouto}{\dot{N}_{\textrm{out,o}}}

\appendix

\section{Linearized optomechanical system with blue-detuned optical pump}

In this section we derive the semi-classical theory for the linear response of an optomechanical crystal pumped on the blue side. 

Following the derivation in Ref.~\cite{safavi2011electromagnetically}, we start by substituting the operators with relevant amplitudes,
\begin{eqnarray}
a &\rightarrow & \alpha_{0} e^{-i \omegap t} + \alpham e^{-i( \omegap + \omega) t} + \alphap e^{-i (\omegap - \omega)t},\\
b & \rightarrow & \beta_{0} + \betam e^{-i\omega t},
\end{eqnarray}
where $\omegap$ is the optical pump frequency and $\omega$ is the microwave drive frequency, either to the EOM or to the IDT. For weak sideband amplitudes $\alphapm \ll \alpha_{0}$, $\alpha_{0}$ is given by
\begin{equation}
\alpha_0 = \frac{-\sqrt{\kappae} \alpha_{\textrm{in,p} } }{ i\Delta + \kappa/2 },
\end{equation}
where $\alpha_{\textrm{in,p}}$ is the input optical pump amplitude. The resulting equations of motion for the sidebands are
\begin{eqnarray}
\pm i \omega \alphapm &=& - (i \Delta + \kappa/2)\alphapm - ig \alpha_0 \betapm - \sqrt{\kappae} \alphainpm, \\
- i\omega \betam &= & - (i \omegam + \gamma/2)\betam - ig (\alpha^*_0 \alpham + \alpha_0 \alphap^*) - \sqrt{\gammae} \betainm, \\
\betap &\equiv& \betam^*.
\end{eqnarray}

When we pump the optical mode at the blue side with $\Delta \equiv \omegac - \omegap \sim - \omegam$ and drive at frequency $\omega \sim \omegam$, the up-converted sideband $\alpham$ is negligible for a sideband-resolved system. For microwave-to-optical (optical-to-microwave conversion) process, $\alphap$ and $\betam$ are solved as a function of input $\betainm$ ($ \alphainp$). The converted output field is given by $ \alphaoutp = \sqrt{\kappae} \alphap $ ($ \betaoutm = \sqrt{\gammae}\betam $).

For optical sideband input and readout ($ \betainm = 0,  \alphaoutp = \alphainp + \sqrt{\kappae} \alphap  $), we derive
\begin{equation}
\Soo(\omega) \equiv \frac{\alphaoutp}{\alphainp} = \frac{-\kappae}{i(\Delta + \omega) + \frac{\kappa}{2} - \frac{G^2}{i(\omega - \omegam) +\gamma/2} },
\end{equation}
where $G = g_{0} |\alpha_0| = g_{0} \sqrt{\nc}$ is the effective optomechanical coupling rate. $\nc\equiv |\alpha_0|^2$ is the intracavity pump photon number. $\Soo$ is directly measured by the VNA and normalized by a background taken with the pump laser far-detuned from the optical mode to remove the response from electronic components. We extract the pump detuning $\Delta$, optical mode decay rates $\kappa$ and $ \kappae$ from $\Soo$. One example of the measured $\Soo$ is shown in Sec.~\ref{SI-sec:optical-sideband-sweep}.

Similarly, the microwave-to-optical and optical-to-microwave conversion scattering parameters can be solved,
\begin{eqnarray}
\Soe \equiv \frac{\alphaoutp}{\betainp} = \frac{ \sqrt{\kappae \gammae} i g_0 \alpha_0}{ [i(\Delta + \omega) + \frac{\kappa}{2}][i(\omega - \omegam) + \frac{\gamma}{2}] - G^2},\\
\Seo \equiv \frac{\betaoutp}{\alphainp} = \frac{ -\sqrt{\kappae \gammae} i g_0 \alpha_0^*}{ [i(\Delta + \omega) + \frac{\kappa}{2}][i(\omega - \omegam) + \frac{\gamma}{2}] - G^2}.
\end{eqnarray}
When the laser is locked to detuning $ \Delta = -\omegam $ and the input frequency is at $\omega = \omegam$, the conversion efficiencies are simplified to
\begin{eqnarray}
\etaoe &=& \etaeo = |\Soe|^2 = |\Seo|^2\\
&=&\frac{\gammae }{\gamma }\frac{ \kappae}{ \kappa} \frac{4C}{(1-C)^2},
\end{eqnarray}
where $C = 4G^2/(\kappa \gamma)$ is the cooperativity. In contrast to the efficiency with red-side-pump $ \eta = \frac{\gammae }{\gamma } \frac{ \kappae}{ \kappa} \frac{4C}{(1+C)^2} $, the blue-side-pump efficiency blows up at $C=1$. Phonon lasing results when $C \ge 1 $ is achieved and the linear theory no longer applies to the system.

In the low cooperativity limit where $C\ll 1$, the conversion efficiency is approximately linear with respect to cooperativity and thus linear with respect to $\nc$,
\begin{equation}
\etaoe = \etaeo = 4C\frac{\gammae }{\gamma }\frac{ \kappae}{ \kappa} = 4 \frac{4g_0^2 \gammae  \kappae }{\gamma^2 \kappa^2} \nc .
\end{equation}

For microwave input and output, we ignore the detailed IDT response for simplicity and combine the IDT mismatch, IDT transduction efficiency and the coupling between the mechanical waveguide and the local mechanical mode to a single parameter $\gammamu$, defined as the mechanical decay rate from the local mechanical mode to the microwave transmission line. In this case, $\beta_{\textrm{in},\pm}$ can be treated as microwave input and output amplitudes, and $\gammae$ is replaced by $\gammamu$. This approximation is valid in the low microwave-to-mechanical conversion efficiency regime. A full analysis of a coupled three-mode system shows that the peak conversion efficiency is given by~\cite{Jiang2019Lithium, hill2012coherent,fang2016optical}
\begin{equation}
\eta = \frac{\kappae}{\kappa} \frac{\kappa_{\textrm{c,e}}}{\kappa_{\textrm{c}}} \frac{4\Cab\Cbc}{(1 \pm \Cab + \Cbc)^2},
\end{equation}
where $\Cab \equiv C$ is the cooperativity between mode $a$ and $b$, and $\Cbc \equiv 4 g_{\textrm{bc}}^2/(\kappa_{\textrm{c}}\gamma) $ is the cooperativity between OMC mechanical mode $b$ and IDT electromechanical mode $c$. $g_{\textrm{bc}} $ is the coupling rate between mode $b$ and $c$. $\kappa_{\textrm{c}}$ and $ \kappa_{\textrm{c,e}} $ are the total and external decay rate of mode $c$ respectively. For weak coupling between $b$ and $c$ where $\Cbc \ll 1$, the efficiency can be approximated by
\begin{equation}
\eta \approx \frac{\kappae}{\kappa}  \frac{4 g_{\textrm{bc}}^2\kappa_{\textrm{c,e}}/\kappa_{\textrm{c}}^2}{\gamma} \frac{4\Cab}{(1 \pm \Cab)^2}.
\end{equation}
By defining $\gammamu \equiv 4 g_{\textrm{bc}}^2 \kappa_{\textrm{c,e}}/\kappa_{\textrm{c}}^2 $, we obtain the two-mode conversion situation with an effective decay rate $\gammamu$ from the mechanical mode to the microwave transmission line. For $\kappa_{\textrm{c,e}}/\kappa_{\textrm{c}}\sim 1$, the weak-coupling condition $ \Cbc \ll 1$ translates to $\gammamu/\gamma \ll 1$, which is valid for the transducer in this work.

In the above derivation, a frequency mismatch $\Delta_{\textrm{bc}}$ between the OMC mechanical mode and the IDT electromechanical mode is not considered. A non-zero $ \Delta_{\textrm{bc}}$ modifies $ \gammamu$ as
\begin{equation}
\gammamu(\Delta_{\textrm{bc}}) = \frac{4 g_{\textrm{bc}}^2 \kappa_{\textrm{c,e}}}{4\Delta_{\textrm{bc}}^2 + \kappa_{\textrm{c}}^2 } = \frac{1}{1 + 4\Delta_{\textrm{bc}}^2/\kappa_{\textrm{c}}^2  } \gammamu(0).
\end{equation}

\section{Energy consumption for converting one classical bit and one quantum bit}

\subsection{Encoding of classical information and relevant energy consumption}

Consider encoding one bit of information onto an optical field via the optomechanical interaction where the frequency of the optical cavity is shifted by $g_0$ with the zero-point motion $x_{\textrm{zp}}$ of the mechanical mode. When the mechanical motion is much slower than the dynamics of the optical cavity ($ \omegamu \ll \kappa $), the optomechanical modulation of the cavity frequency can be approximated as quasi-static. When a coherent state is injected and reflected from the optical cavity, a displacement that is approximately $ x_{\pi} = x_{\textrm{zp}}\kappa/g_0 $ gives a significant phase shift~\cite{Safavi-Naeini2019}. The corresponding mechanical energy is
\begin{equation}
E_{\textrm{mech}} = \frac{1}{2}m_{\textrm{eff}}\omegam^2 x^2_{\pi} = \frac{1}{2}m_{\textrm{eff}}\omegam^2 x^2_{\textrm{zp}} \frac{\kappa^2}{g_0^2} = \frac{\hbar \omegam \kappa^2}{4g_0^2}.
\end{equation}
For an imperfect conversion between microwave and mechanical energy, the energy-per-bit is given by
\begin{equation}
\label{eqn:SI-Ebit}
E_{\textrm{bit}} = \frac{E_{\textrm{mech}}}{\etam} =  \hbar \omegam \frac{ \kappa^2}{4g_0^2} \frac{1}{\etam}.
\end{equation}

In the sideband resolved regime where $ \omegam \gtrsim \kappa$, the dynamics of the optical cavity has to be considered. For a coherent state  $ \alpha_0 $ that is initially in the cavity, the classical equation of motion is
\begin{equation}
\dot \alpha = -\left( \frac{\kappa}{2} + ig_0 \sqrt{\nphon} \cos (\omegam t+\phi) \right) \alpha,
\end{equation}
where we have chosen the frame rotating at the static cavity frequency $\omegac$. The information can be encoded in the phase factor with $\phi = 0$ or $\phi = \pi$. We choose this encoding method because it returns to a simple phase shift in the optical field in the quasi-static limit $\omegam \ll \kappa$. The time-dependent solution of the equation of motion is
\begin{equation}
\alpha (t) =  \alpha_0 \exp \left(- \frac{\kappa t}{2} - i h \sin (\omegam t+\phi) \right).
\end{equation}
We have defined the modulation index $ h \equiv g_0 \sqrt{\nphon}/\omegam$. The output field is given by $ \alpha_{\textrm{out}}(t) = \sqrt{\kappa} \alpha(t) $. We set $ \kappae = \kappa $ for simplicity. 

In a completely classical theory of electromagnetics, an arbitrarily small change in the optical field can be in principle detected deterministically with a sufficiently sensitive measurement. Quantum noise limits imprecision of estimating observables such as the phase of the an optical field. For example the imprecision in phase of coherent states is given by the standard quantum limit (SQL) $\Delta \phi~\approx 1/\sqrt{n}$, where $n$ is the average number of photons in the field~\cite{clerk2010introduction}. Therefore, to determine whether a device has sufficiently changed the state of the field to encode a bit, we need to consider how an initial coherent state $|\alpha\rangle$ with fixed optical energy (here we take $n=|\alpha_0|^2=1$) is modified by the device and use the quantum detection theory~\cite{helstrom1969quantum} to calculate the probability of error $P_e$ in distinguishing resulting states $|\Psi_k\rangle$. Note that this error rate is absolutely the lowest that can be achieved given with any possible receiver. Since our optomechanical device is an open quantum system~\cite{wiseman2009quantum} where the field leaks out into a waveguide, we consider the output states to be in the Hilbert state of the all of the waveguide states, which is generated  by an operator that is a superposition of the time-dependent output operators with a time-domain waveform that corresponds to the classical solution,
\begin{equation}
\opA \equiv \int_0^\infty dt  f^*(t) \aout (t).
\end{equation}
Here  $ f(t) =\sqrt{\kappa} \exp \left(-\kappa t/2 - i h \sin (\omegam t+\phi) \right)  $, and $\aout(t)$ is the annihilation operator for each time bin that obeys $ [\aout(t), \aout(t')]=0$ and $[\aout (t), \aout^\dagger (t')] = \delta(t-t') $. It is straightforward to verify that $\opA $ is a properly normalized bosonic operator such that $[\opA, \opA]=0$ and $[\opA, \opAd] = 1 $. Note that the information is encoded in $\opA$ via the phase factor $\phi$ in $f(t)$. From now on we explicitly denote the resulting operator with suffix as $\opA_\phi $, where  $ \phi = 0 $ or $\phi = \pi$. The commutation relations between operators with different values of $\phi$ are now $ [\opA_\phi, \opA_{\phi'}]=0 $ and
\begin{equation}
\label{eqn:SI-A-commutator}
[\opA_\phi, \opAd_{\phi'}] = \int_0^\infty  f_\phi^*(t)f_{\phi'}(t) dt = \int_0^\infty dt \kappa e^{-\kappa t}e^{ih[\sin(\omegam t + \phi) - \sin(\omegam t + \phi')] }.
\end{equation}

For an initial coherent state $ \ket{\alpha_0} $ in the cavity, the output state is given by the coherent state generated by the operator $A_\phi$,
\begin{equation}
\ket{\Psi_\phi} = e^{-|\alpha_0|^2/2} e^{\alpha_0 \opAd_\phi} \ket{0}.
\end{equation}
We use $\ket 0 $ to represent the vacuum state in the continuous time domain.

A criterion is required to determine whether a bit of information is successfully encoded or not. It is naturally related to how separate are the two output state $\ket{\Psi_0}$ and $\ket{\Psi_\pi}$, and how well we could distinguish them by measurement which can be characterized by an error possibility $ P_\textrm{e} $ of misidentifying the state. The error possibility is bounded below by the Helstrom-Holevo lower bound~\cite{wiseman2009quantum}. For distinguishing two pure states with equal prior probability, the Helstrom-Holevo lower bound is given by
\begin{equation}
P_\textrm{e} = \frac{1}{2} \left( 1 - \sqrt{1 - F} \right),
\end{equation}
where $ F \equiv |\braket{\Psi_0 | \Psi_\pi}|^2 $ is the fidelity between the two encoded states,
\begin{eqnarray}
\label{eqn:SI-fidelity}
F & = & \left|e^{-|\alpha_0|^2}\bra{0} e^{\alpha_0^* \opA_0} e^{\alpha_0 \opA_\pi} \ket{0} \right|^2 \notag \\
& = & \left|\exp{\left[-|\alpha_0|^2(1-[ \opA_0, \opAd_\pi])\right]}\right|^2.
\end{eqnarray}
Intuitively, $ P_\textrm{e} = 0 $ for two orthogonal state with $ F = 0 $ and $ P_\textrm{e} = 1/2 $ for two identical state with $F=1$. A low fidelity between the two states is desired for them to be more separable. For $F\ll 1 $, $ P_\textrm{e} \approx F/4 $. In addition, when a larger coherent state $\alpha_0$ is used, the more separate the two states are for fixed $\nphon$ and $h$. It is important to consider the optical field used for the encoding, since a stronger field makes it easier to measure a smaller change in its properties, reducing the energy required to modify the optical material that is used for imposing the change.

\subsubsection{Slow limit: $\omegam \ll \kappa $}

Before we proceed with numerical calculation, it is instructive to look at the quasi-static or the slow-limit where $ \omegam \ll \kappa $. In this situation, the difference between operators $\opA_0 $ and $\opA_\pi$ is approximately a phase shift $ \Delta \phi = \pm i h \omegam t $. The commutator and the fidelity are simplified to
\begin{eqnarray}
[ \opA_0, \opAd_\pi] &=& \int_0^\infty dt \kappa e^{-\kappa t}e^{2ih\sin\omegam t } \approx \frac{\kappa}{\kappa - 2ih\omegam},\\
F &\approx& \exp \left(-2|\alpha_0|^2 \frac{h^2\omegam^2}{h^2\omegam^2 + (\kappa/2)^2}\right) = \exp \left(-2|\alpha_0|^2 \frac{g_0^2 \nphon}{g_0^2 \nphon + (\kappa/2)^2}\right).
\end{eqnarray}
As a result, the fidelity is limited by $ F_{\textrm{min}} = \exp(-2|\alpha_0|^2) $ and the characteristic modulation index for achieving a low fidelity and a low error probability occurs at $ g_0 \sqrt{\nphon} = \kappa/2 $. The corresponding $F \approx \exp (-|\alpha_0|^2) $, $ \nphon = \kappa^2/(4g_0^2) $ and
\begin{equation}
E_{\textrm{bit,slow}} = \frac{\hbar \omegam \nphon} {\etam} = \hbar \omegam \frac{ \kappa^2}{4g_0^2} \frac{1}{\etam},
\end{equation}
which is identical to Eq.~\ref{eqn:SI-Ebit}. For an initial coherent state with an average photon number $|\alpha_0|^2 = 1$, we calculate the fidelity $F = 1/e$ and the error probability $P_\textrm{e} = 10.25\% $.

\subsubsection{Fast limit: $\omegam \gg \kappa $}

Now we consider the fast-limit where $ \omegam \gg \kappa $. By expanding the sinusoidal phase modulation into sidebands, the commutator and the fidelity are approximated by
\begin{eqnarray}
[ \opA_0, \opAd_\pi] &=& \int_0^\infty dt \kappa e^{-\kappa t} \sum_n J_n(2h)e^{in\omegam t} = \sum_n J_n(2h) \frac{\kappa}{\kappa - in\omegam} \approx J_0(2h),\\
F& \approx & \exp\left(-2|\alpha_0|^2 (1-J_0(2h)) \right).
\end{eqnarray}
$J_n$ is the Bessel function of the first kind. The minimal fidelity and minimal error probability are achieved at the first minimal of $ J_0(2h) \approx -0.4$, where $ 2h \approx 3.832$. Comparing to the slow-limit situation, the fast-limit fidelity is much lower at this optimal modulation index. To have a better comparison among the slow-limit, fast-limit and the general situation, we look for the modulation index where $J_0(2h) \sim 1/2 $ instead of the optimal modulation index, which gives a fidelity $F \approx  \exp (-|\alpha_0|^2)  $, similar to the slow-limit fidelity. The corresponding $h \approx 0.76$, $ \nphon =  h^2\omegam^2/g_0^2 \approx \omegam^2/(2g_0^2) $ and
\begin{equation}
E_{\textrm{bit,fast}} = \frac{\hbar \omegam \nphon} {\etam} = \hbar \omegam \frac{ \omegam^2}{2g_0^2} \frac{1}{\etam}. 
\end{equation}

\subsubsection{Numerical calculation for the general situation}

In the general case, we assume that the best measurement is adopted to achieve the Helstrom-Holevo lower bound and we look at the modulation index $h$ or equivalently the intracavity phonon number $\nphon$ that is required to reach a given error rate. We evaluate Eq.~\ref{eqn:SI-fidelity} and Eq.~\ref{eqn:SI-A-commutator} numerically for different ratio between $ \omegam $ and $\kappa$, and we search for the $\nphon$ required to reach $ P_\textrm{e} = 10\% $ for an initial coherent state with an average photon number $|\alpha_0|^2 = 1 $.

\begin{figure}[htbp]
	\centering
	\includegraphics[scale=0.5]{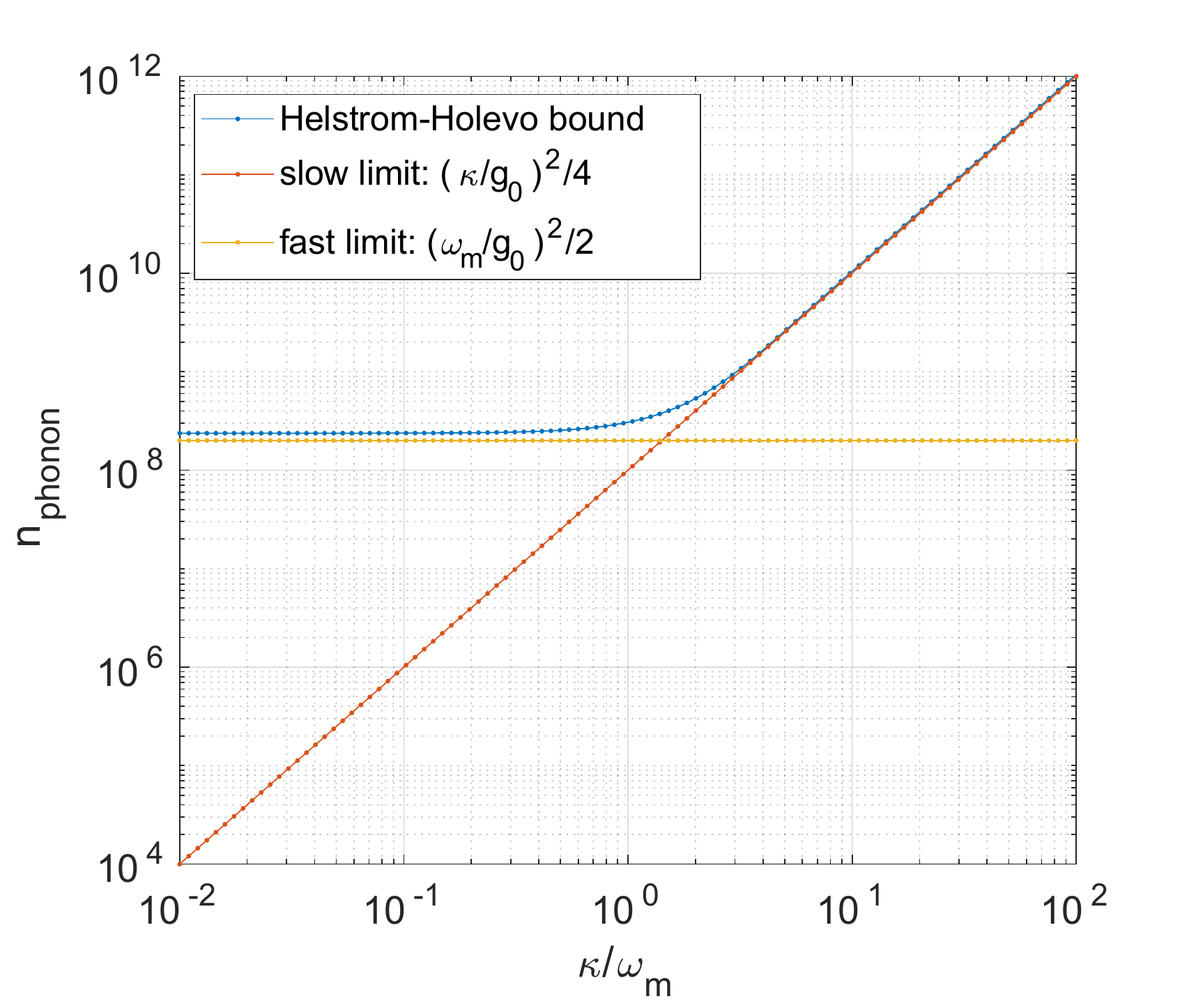}
	\caption{\label{SIFig:Ebit} Number of intracavity phonons required to achieve an error probability $ P_{\textrm{e}}=10\% $ for distinguishing between two encoded states with an initial coherent state which has an average photon number $ |\alpha_0|^2 = 1 $ in the optical cavity. Blue line is the numerical calculation from Helstrom-Holevo bound. Red (yellow) line is the approximated result in the slow (fast) limit where $\omegam \ll \kappa $ ($\omegam \gg \kappa $).}
\end{figure}

We show the calculated $\nphon $ in Fig.~\ref{SIFig:Ebit} for fixed $g_0$ and $\omegam $, and varying $ \kappa $ by four orders of magnitude. We adopt $ g_0/\omegam = 5\times 10^{-5} $ for the calculation, which closely represents the device demonstrated in this work. We clearly observe that $\nphon $ given by the Helstrom-Holevo bound is well approximated by the slow and fast limit in the corresponding regimes. Interestingly, for a sideband-unresolved piezo-optomechanical transducer device with a given mechanical mode frequency $\omegam$ and an optomechanical coupling $g_0$, decreasing the optical linewidth lowers the energy per bit until $ \kappa \sim \omegam $. Further increasing the optical quality factor no longer helps for reducing the energy consumption.

\subsection{Energy consumption in quantum frequency conversion}

In the case of quantum transduction, an unity internal conversion efficiency is desired. For low microwave-to-mechanical conversion efficiency, best internal conversion is achieved at optomechanical cooperativity $C \equiv 4g_0^2\nc/(\kappa\gamma) = 1$. The non-zero optical cavity internal decay rate $\kappai$ leads to a power dissipation
\begin{equation}
P_{\textrm{diss}} = \hbar \omegac \nc \kappai.
\end{equation}
The conversion rate is limited by the mechanical-to-microwave decay rate $\gammamu$, leading to the energy-per-qubit
\begin{equation}
\frac{P_{\textrm{diss}}}{\gammamu} = \frac{\hbar \omegac \kappa\kappai}{4g_0^2 \etam}.
\end{equation}
However, the converted photon only successfully leaves the converter at efficiency $\etao \equiv \kappae/\kappa$, adding another factor to the actual energy-per-qubit,
\begin{equation}
E_{\textrm{qubit}} = \hbar \omegac \frac{ \kappa\kappai}{4g_0^2 } \frac{1}{\etam\etao}.
\end{equation}

\section{Mode decomposition of the IDT and the OMC leakage in the mechanical waveguide}

We showed the simulated IDT and OMC modes in the mechanical waveguide region in the main text. Here we further perform mode decomposition of the leakage mode to the guided modes of the mechanical waveguide~\cite{dahmani2019piezoelectric}. The resulting fractional powers in different waveguide modes are listed in Table~\ref{tab:mode-decomp}. The OMC leakage motion is decomposed after the curved waveguide.

We list seven most relevant waveguide modes in the table, including the first-order longitudinal (L1), first-order horizontal-shear (SH1), second-order Lamb (A2), fundamental horizontal-shear (SH0), first-order Lamb (A1), fundamental Lamb (A0) and fundamental longitudinal (L0). The higher order modes have one or more nodes along the in-plane transverse direction, because the waveguide width is much larger than the thickness.

\begin{figure*}[tb]
	\centering
	\includegraphics[scale=0.6]{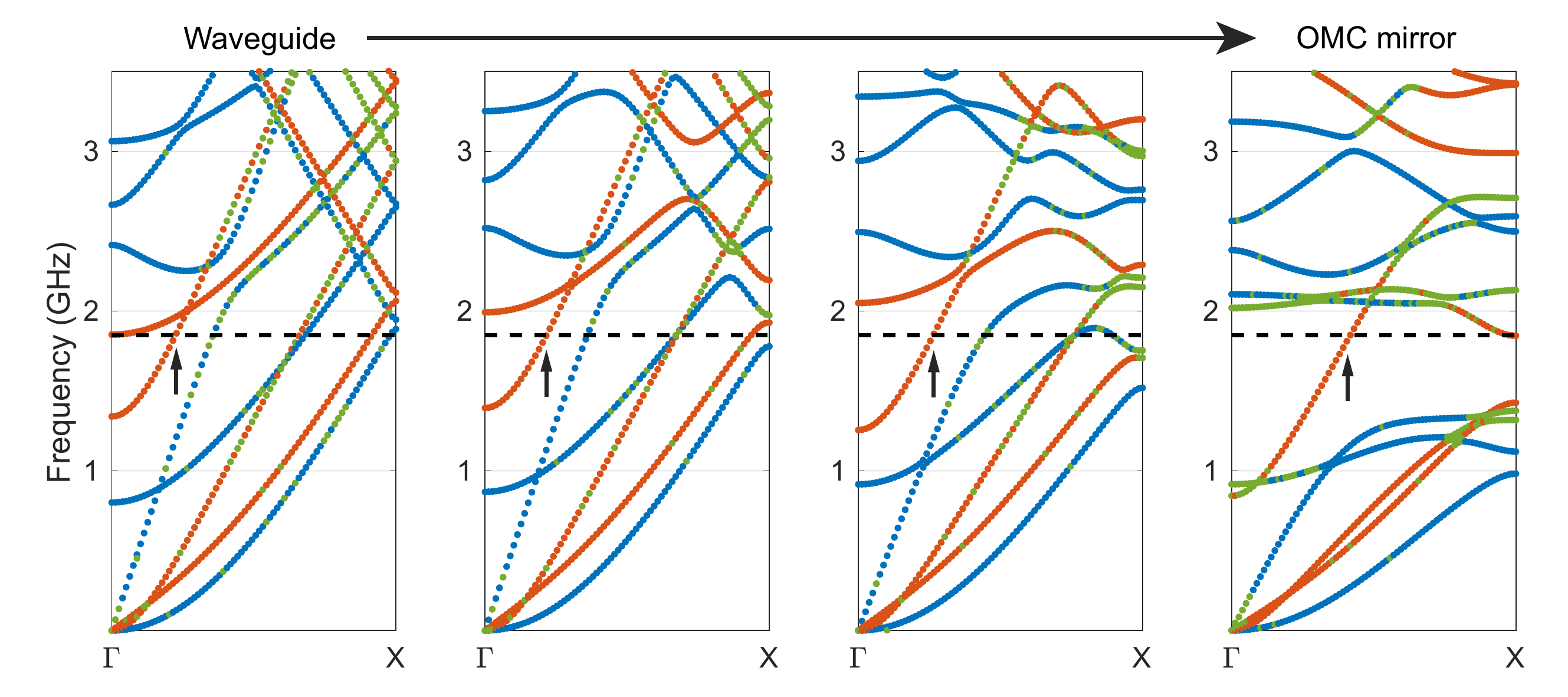}
	\caption{\label{SIFig:WG2OMC-bands} Evolution of mechanical bands of the linear taper from the simple waveguide to the OMC mirror cell. Sub-figures from left to right correspond to band structures of the first taper unitcell with minor geometric change (left), the OMC mirror cell band structure (right) and two intermediate unitcells between the waveguide and the OMC mirror cell geometry. The bands are classified by semi-y-symmetric (blue), semi-y-asymmetric (red) and mixed (green). Horizontal dashed lines represent the OMC breathing mode frequency. Black arrows indicate the first-order longitudinal motion band.}
\end{figure*}

Intuitively, the mechanical modes supported by the OMC mirror cell near the local breathing mode frequency are mostly asymmetric along the nanobeam symmetry plane. When the crystal mirror symmetry plane is aligned to the OMC mirror symmetry plane, the mirror cell modes near the OMC breathing mode frequency are strictly asymmetric~\cite{Jiang2019Lithium}. As a result, the OMC mechanical leakage mostly enters the asymmetric waveguide modes such as L1 and SH1. The L1 mode has the smallest wave-vector, making it more strongly coupled to the Gamma-point OMC breathing mode. This is further illustrated in Fig.~\ref{SIFig:WG2OMC-bands}, where the evolution of the mechanical band structure from the waveguide geometry to the OMC mirror cell geometry is shown as four snapshots. There is no crystal  symmetry along the geometric y-symmetry plane, but we could still classify the modes as semi-symmetric (blue), semi-asymmetric (red) or mixed (green) by evaluating proper overlap integrals. We observe that the L1 band always covers the OMC breathing mode frequency, indicated by the horizontal dashed lines and the black arrows.

\begin{table}
	\caption{Mode decomposition of the IDT and OMC leakage mode to the mechanical waveguide modes. Numbers in percentage are the fractional power in corresponding modes.} \label{tab:mode-decomp}
	\begin{center}
		\begin{tabular}{c|c|ccc}
			\hline
			Waveguide modes  & OMC$\rightarrow$WG & IDT$\rightarrow$WG & IDT $\eta $ & $T_{\textrm{b}\upmu} $\\
			\hline
			L1& $64.1\% $ &$74.6\%$  & $0.6\% $ & $ 0.19\% $ \\
			SH1  & $ 28.0\% $ &$<0.01\%$ & $0$ & $0 $ \\
			A2 & $ 5.4\% $& $4.37\%$ & $0.04\% $  & $ 0.013 \% $ \\
			SH0  & $ 2.0\% $& $ <0.01\%$ & $0 $& $0 $ \\
			A1  & $ 0.1\% $& $ 17.9\% $ & $0.14\% $ & $0.044\% $ \\
			A0  & $ 0.2\% $ & $ 0.4\%$ & $ <0.01\% $&  $0 $  \\
			L0  &  $< 0.1\% $ & $2.7\% $ & $0.02\% $   &$ 0.006\% $\\
			\hline
		\end{tabular}
	\end{center}
\end{table}

In the IDT simulation, a material loss tangent that corresponds to a quality factor $ Q_{\textrm{i}} = 800 $ is added to roughly match the simulated and measured peak conductance. We further define the IDT efficiency $\eta$ in terms of the fractional power in different waveguide modes over the total power dissipated in the IDT, including material loss and clamping loss. We list the simulated IDT $\eta $ in Table.~\ref{tab:mode-decomp} and deduce that $99.2\%$ of the total absorbed microwave power is dissipated in material loss and clamping. The material loss can be drastically eliminated by going to cryogenic temperature, while the clamping loss can be minimized by deploying phononic shield~\cite{alegre2011quasi,maccabe2019phononic}. In the actual device, $ 31.6\% $ of the incident microwave power is absorbed by the transducer, from which we calculate the microwave-to-waveguide-mode scattering parameter $ T_{b\upmu} $. This is the ratio between the outgoing mechanical power in waveguide modes and the incident microwave power.

The measured microwave-to-mechanical conversion efficiency $ \etam \equiv \gammamu/\gamma $ could be higher than the maximal $T_{b\upmu}$ in the table. The OMC mechanical mode is coupled to multiple waveguide modes, and the standing wave resonances of the mechanical waveguide modify the overall response of the IDT-waveguide-OMC system.

\section{Optomechanical backaction measurement}

Here we show the optomechanical backaction measurement of the optomechanical coupling $g_0$ and the mechanical linewidth $\gamma$.

\begin{figure}[htbp]
	\centering
	\includegraphics[scale=0.6]{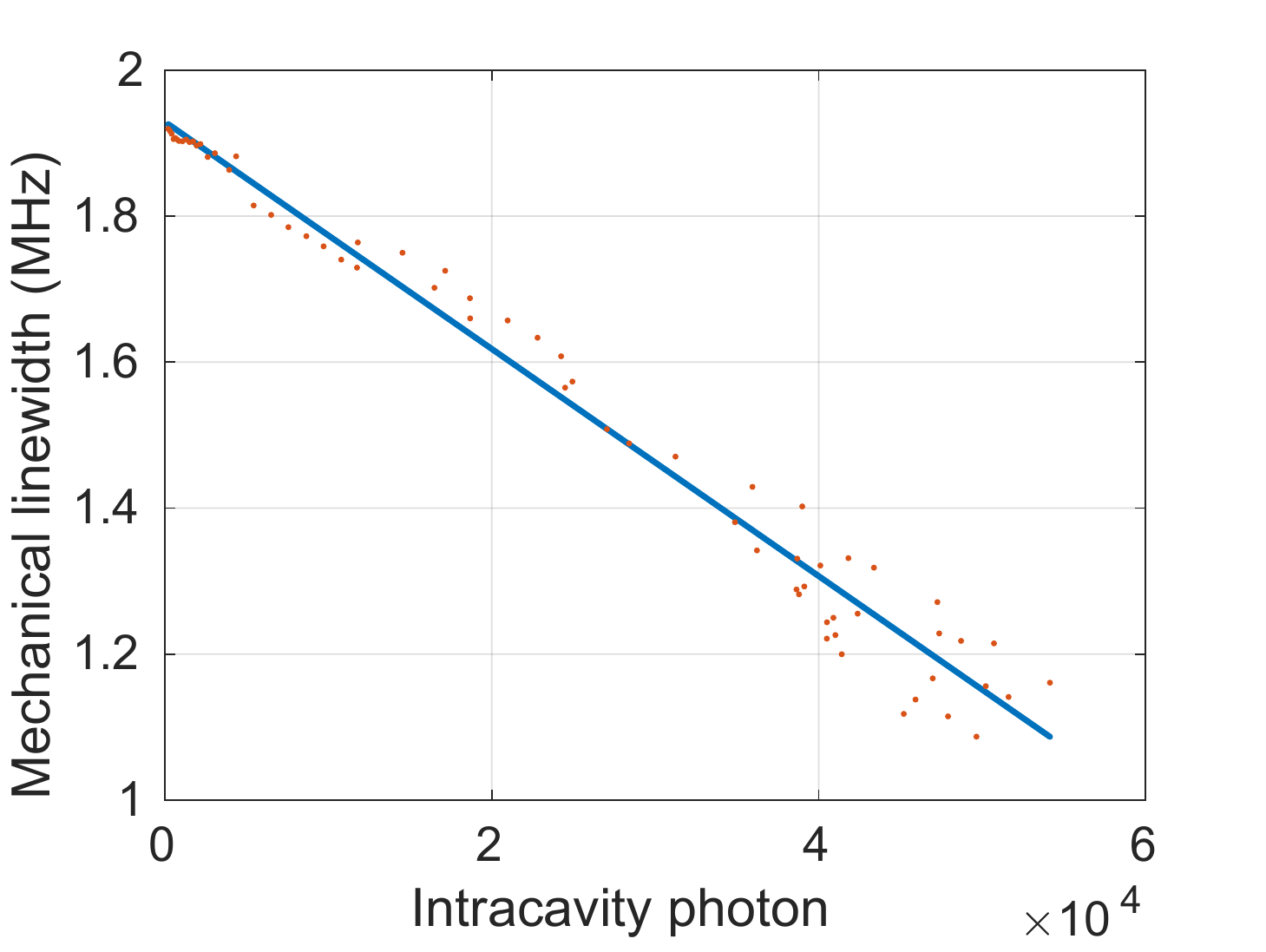}
	\caption{\label{SIFig:OM-backaction-fit} Optomechanical backaction measurements and fit.}
\end{figure}

Figure \ref{SIFig:OM-backaction-fit} shows the extracted mechanical linewidth versus intracavity photon number with blue detuned pump laser at $\Delta = -\omegam$. Blue curve is a linear fit, giving an intrinsic mechanical linewidth of $\gamma/2\pi = 1.93~\textrm{MHz}$ and an optomechanical coupling rate $g_0/2\pi = 70~\textrm{kHz} $.

\section{Extracting pump detuning and optical cavity decay rates}
\label{SI-sec:optical-sideband-sweep}

\begin{figure}[t]
	\centering
	\includegraphics[scale=0.6]{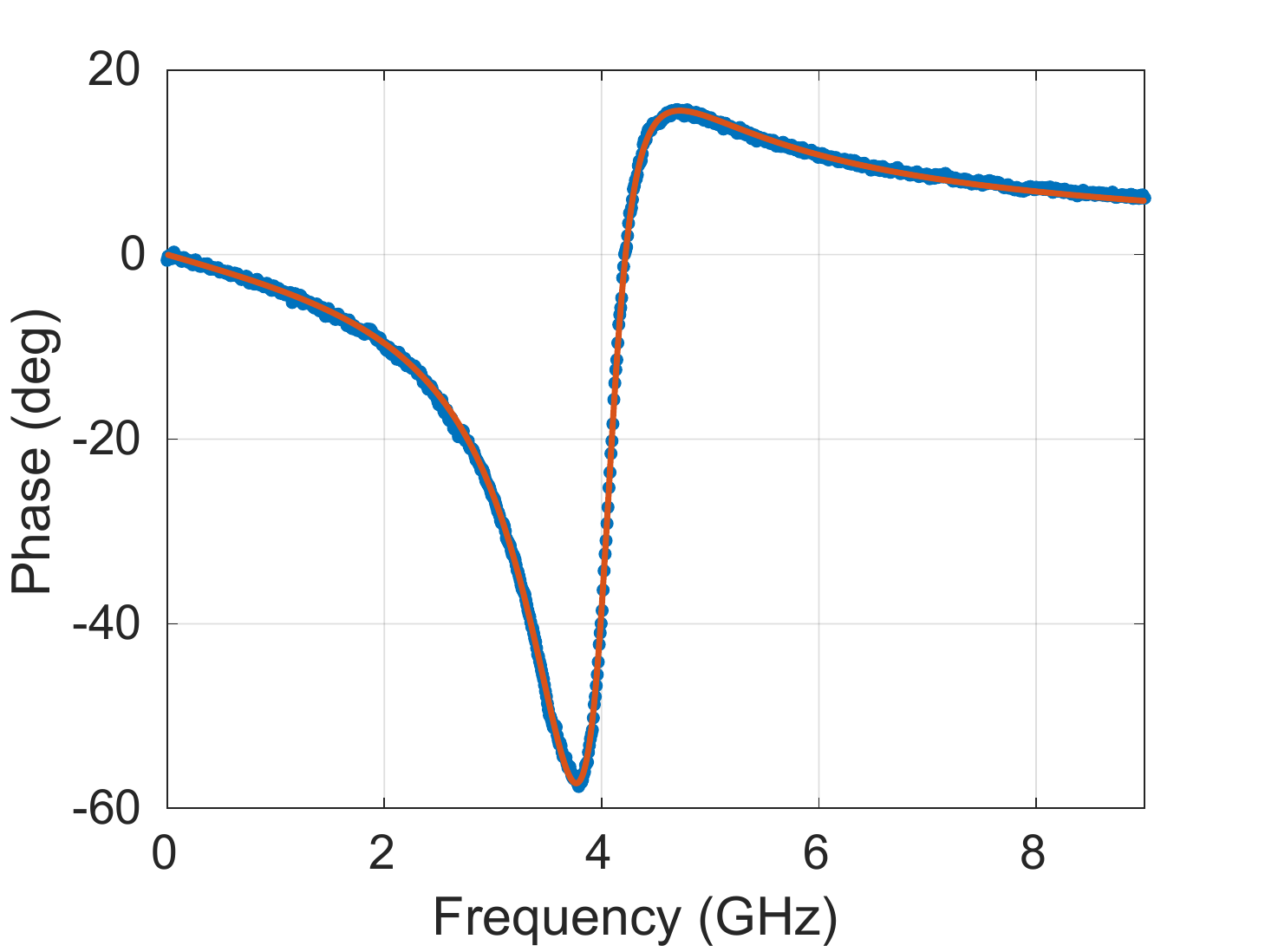}
	\caption{\label{SIFig:optical-sideband-sweep} Optical sideband response. Blue dots: measured phase response. Red curve: fit.}
\end{figure}

We use the optical sideband response (optical to optical scattering parameter) of the device to extract the cavity-laser detuning $\Delta$ and the optical cavity decay rates $\kappa$ and $\kappae$. Figure~\ref{SIFig:optical-sideband-sweep} shows a typical phase response of the optical-to-optical $S$ parameter. A fit curve based on the linearized optomechanical response theory is also shown as the red curve.

For this specific measurement, we extract a detuning $\Delta/2\pi = -3.698~\textrm{GHz}$, $\kappa/2\pi = 1203~\textrm{MHz}$ and $\kappae/2\pi = 781~\textrm{MHz}$. A significant phase response is the signature of an over-coupled optical mode where $\kappae > \kappa/2$. A clear deviation from the measurement is observed if the fit curve is forced to have $\kappae < \kappa/2$.

\section{Extracting the acousto-optic modulation index and $ V_\pi $}

In this section we derive the optical cavity spectrum under acousto-optic modulation and show typical fit results for extracting the acousto-optic modulation index $ h $.

When the optical cavity's resonant frequency $\omegac$ is being modulated, in the frame where the cavity frequency is fixed, the laser input is phase-modulated by the same strength and splits into sidebands with relative amplitudes given by Bessel functions of the first kind. We expect the output spectrum to be a convolution between the phase modulation combs and the Lorentzian cavity response. We derive the rigorous output spectrum in the following.

We start with the classical equation of motion for the optical field amplitude in the rotating frame of the laser,
\begin{equation}
\dot{\alpha} = -(i \Delta + \frac{\kappa}{2})\alpha - i g_0 \sqrt{\nphon} \cos (\omegamu t)\alpha - \sqrt{\kappae} \alphain,
\end{equation}
where $g_0$ is the zero-point optomechanical coupling rate and $\nphon$ is the intracavity phonon number from the microwave drive at frequency $\omegamu $. We use the following substitution to eliminate the time-dependent term on the right hand side:
\begin{equation}
\alpha \rightarrow \alpha' \exp [ -i h  \sin (\omegamu t) ].
\end{equation}
We have introduced the acousto-optic modulation index $h \equiv g_0 \sqrt{\nphon}/\omegamu$. After the substitution, the equation for $\alpha'$ is
\begin{eqnarray}
\dot \alpha' &=& -(i \Delta + \frac{\kappa}{2})\alpha'  - \sqrt{\kappae} \alphain \exp(i h \sin \omegamu t )\\
&=&-(i \Delta + \frac{\kappa}{2})\alpha'- \sqrt{\kappae} \alphain \sum_n J_n (h) e^{in\omegamu t},
\end{eqnarray}
where $ J_n (h) $ are Bessel functions of the first kind. By separating $\alpha'$ further into sidebands $\alpha' = \sum_n \alpha'_n \exp(i n \omegamu t)$, $\alpha'_n$ is time-independent and is given by
\begin{equation}
\alpha'_n = \frac{-\sqrt{\kappae} \alphain J_n(h)}{i(\Delta + n \omegamu) + \kappa/2}.
\end{equation}

The output optical field is
\begin{eqnarray}
\alphaout &=& \alphain +\sqrt{\kappae} \exp(-i h \sin \omegamu t) \sum_n \alpha'_n e^{in\omegamu t} \\
&=&\alphain \exp(-i h \sin \omegamu t)\cdot \notag\\
& & \left( \sum_n J_n(h)e^{in\omegamu t}   - \kappae \sum_{n} \frac{ J_n(h)  e^{in\omegamu t}}{i(\Delta + n \omegamu) + \kappa/2} \right).
\end{eqnarray}
The slow photodetector used for the measurement selects the direct-current component of $ |\alphaout|^2 $. As a result, the normalized reflection is a weighted sum of Lorentzians
\begin{eqnarray}
R &\equiv& \left< \left|\frac{\alphaout}{\alphain}\right|^2 \right> \notag \\
&=& \left<\left|\sum_n J_n(h)e^{in\omegamu t} \left( 1 - \frac{\kappae}{i(\Delta + n \omegamu) + \kappa/2} \right) \right|^2\right> \notag \\
&=& \sum_n J_n(h)^2 \left| 1 - \frac{\kappae}{i(\Delta + n \omegamu) + \kappa/2}  \right|^2  \label{eqn:acousto-optic-spectrum}
\end{eqnarray}

We use Eq.~\ref{eqn:acousto-optic-spectrum} to fit the measured reflection spectrum and extract the acousto-optic modulation index $h $ for different microwave drive powers and frequencies as shown in the main text. We observed acousto-optic sidebands up to $ h \sim 9$, hence the sum for $n$ is truncated at $\pm 20$, far enough comparing to the optical linewidth. The only fit parameters are $h$ and a frequency shift and scaling for compensating the difference between the laser wavelength readout and the microwave frequency. Scaling the frequency axis does not affect the unitless parameter $h$.

\begin{figure}[htbp]
	\centering
	\includegraphics[scale=0.6]{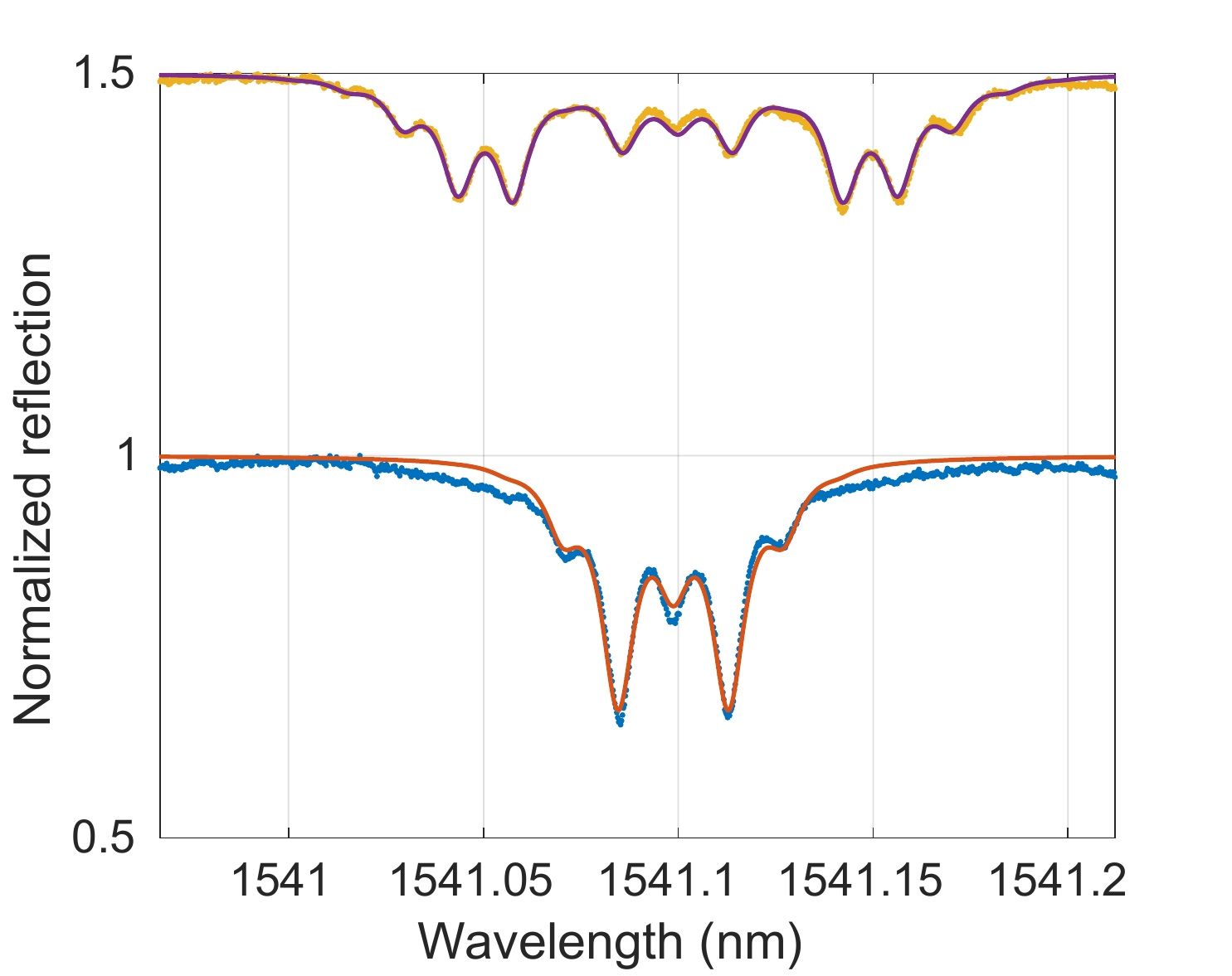}
	\caption{\label{SIFig:AOM-fit} Normalized optical reflection spectrum with acousto-optic modulation and corresponding fit curves.}
\end{figure}

Figure \ref{SIFig:AOM-fit} shows two typical reflection measurements and corresponding fit curves with $h = 1.747$ and $h = 4.812$ (shifted up by 0.5) respectively. Good fitting results are observed up to $h \sim 8$. For $h > 8$, the spectrum spreads out for a wide range of frequency, resulting in smaller reflection dips that are hard to be captured by the fit without a good initial guess.

The measured acousto-optic modulation index $h$ gives us a relation between the OM coupling $g_0$ and the microwave-mechanical external coupling rate $\gammamu$. From standard input-output theory, the intracavity phonon number from input microwave photon flux $\Ndotinmu = P_\upmu/(\hbar\omegamu)$ is
\begin{equation}
\nphon = \frac{\gammamu \Ndotinmu }{ (\omegam - \omegamu)^2 + (\gamma/2)^2}.
\end{equation}
When $\omegamu = \omegam$, the above expression is simplified to $ \nphon = 4\gammamu \Ndotinmu /\gamma^2 $. Combining with the definition of modulation index $h$, we have
\begin{equation}
\gammamu= \frac{h^2 \omegamu^2 \gamma^2}{4g_0^2\Ndotinmu}.
\end{equation}
Underestimation of $g_0$ and overestimation of $h$ and $\gamma$ could lead to a larger $\gammamu$.

Since the extra phase variation of the cavity is given by $ \phi(t) =  h \sin (\omegamu t) $, it is natural to define the voltage required to obtain a $\pi$ phase shift as when $ h_{\pi,\textrm{PM}} \equiv g_0 \sqrt{\nphon}/\omegamu= \pi $ is achieved. The microwave voltage is implicitly included via the microwave power $P_\upmu $ in $\nphon$. Such a voltage is given by
\begin{equation}
V_{\pi,\textrm{PM}} = \frac{h_{\pi,\textrm{PM}} \sqrt{2P_{\upmu}Z_0} }{h},
\end{equation}
where $P_{\upmu}$ is the microwave power used to achieve the measured $h$. $Z_0 = \SI{50}{\ohm}$ is the impedance of the microwave transmission line.

We measured a maximal $h = 3.518$ with $P_{\upmu} = \SI{7.24}{\micro\watt}$. The deduced phase-modulation $V_{\pi,\textrm{PM}}$ is \SI{24.0}{\milli\volt}.

\section{Wide range IDT response and microwave-to-optical conversion measurement}

\begin{figure*}[tb]
	\centering
	\includegraphics[scale=0.45]{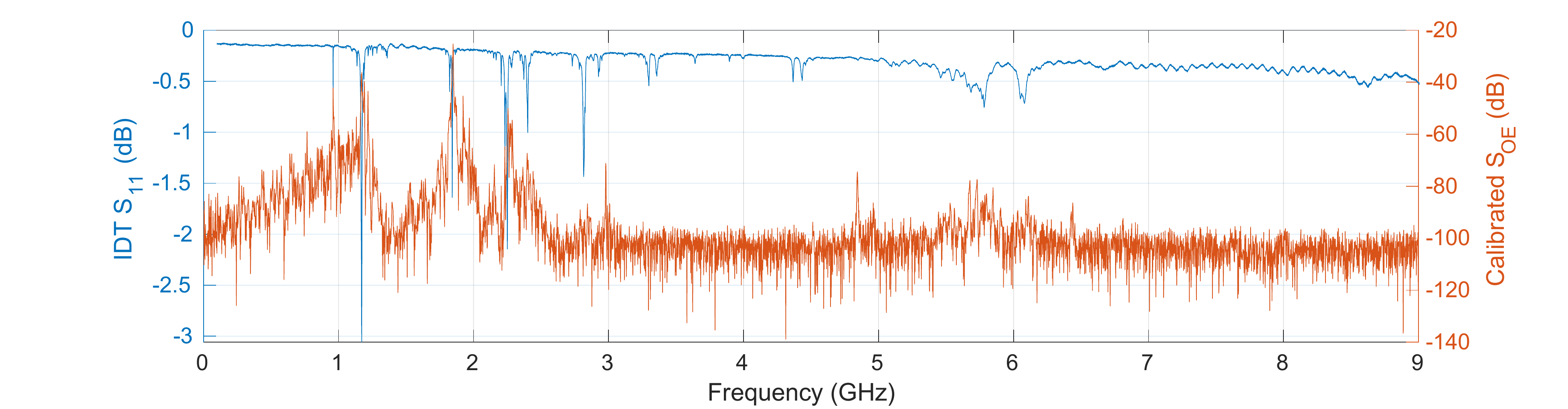}
	\caption{\label{SIFig:WideIDT-and-OE} Wide range data for IDT $S_{11}$ and microwave-to-optical $S$ parameter. }
\end{figure*}

We show the wide frequency range IDT $S_{11}$ response in Fig.~\ref{SIFig:WideIDT-and-OE}. The IDT response is measured with identical setup as described in Ref.~\cite{dahmani2019piezoelectric}. A typical microwave-to-optical $\Soe$ parameter from $100~\textrm{kHz}$ to $9~\textrm{GHz}$ is also shown in Fig.~\ref{SIFig:WideIDT-and-OE}. The $\Soe$ parameter is shifted according to the calibrated peak efficiency.

\section{Microwave-to-optical conversion efficiency calibration}

The microwave-to-optical conversion efficiency is defined as
\begin{equation}
\label{eqn:etaoe-def}
\etaoe \equiv \frac{\Ndotouto}{\Ndotinmu},
\end{equation}
where $\Ndotouto$ is the output optical sideband photon flux after the photons enter the lensed fiber, and $\Ndotinmu$ is the input microwave photon flux before the IDT. $\etaoe$ is the total efficiency including the IDT mismatch and transduction efficiency, and the fiber-to-chip optical coupling efficiency $\eta_{\textrm{oc}} \sim 65\%$.

$\Ndotinmu$ is given by
\begin{equation}
\Ndotinmu = \eta_{\textrm{cable}} \Pinmu / (\hbar \omegamu),
\end{equation}
where $\eta_{\textrm{cable}} = 57.5\%$ is the microwave cable loss and $\Pinmu$ is the VNA output power.

$\Ndotouto$ is calibrated using the method introduced in Ref.~\cite{patel2018single} and also briefly described here. After each measurement, the pump laser is detuned from the optical cavity, and a second laser with optical power $P_{\textrm{cal,o}}\sim 150~\textrm{nW} \ll \Pin$ is tuned to $\sim \omegam$ away from the pump laser without changing any other settings in the measurement setup. The beat tone between the pump laser and the calibration laser is measured by the highspeed detector and the RSA. The pump laser power $\Pin$ is then fully attenuated to $\ll 1~\textrm{nW}$, and the calibration laser power $P_{\textrm{cal,o}}$ is measured by a sensitive power meter (PM).

By integrating the microwave power of the beat tone $P_{\textrm{cal},\upmu}$ on the RSA and measure the optical insertion loss $\eta_{\textrm{out}} = 63.6\%$ from the lensed fiber to the power meter, we get the optical detection gain
\begin{equation}
G \equiv \frac{ P_{\textrm{cal},\upmu} }{ P_{\textrm{cal,o}}/\eta_{\textrm{out}} },
\end{equation}
which is between the optical power at the lensed fiber output and the microwave power at the RSA.

We measure the converted optical sideband using the highspeed detector and the RSA with VNA output frequency fixed at the peak conversion frequency. The measured microwave power $P_{\textrm{out},\upmu}$ on RSA is then converted to optical sideband power $P_{\textrm{out,o}}$ at the lens fiber output using the calibrated detection gain $G$. The converted sideband photon flux is then calculated by
\begin{equation}
\Ndotouto =\frac{P_{\textrm{out,o}}}{\hbar\omegac}= \frac{P_{\textrm{out},\upmu}}{G\hbar\omegac}.
\end{equation}

\section{Optical-to-microwave conversion efficiency calibration}
\label{sec:O2E-calib}

The optical-to-microwave conversion efficiency is defined similar to Eq.~\ref{eqn:etaoe-def} as
\begin{equation}
\etaeo \equiv \frac{\Ndotoutmu}{\Ndotino}.
\end{equation}
$\Ndotino$ is the input optical sideband photon flux before the lensed fiber and $\Ndotoutmu$ is the converted microwave photon flux at the IDT output.

Given the measured $S_{21}$ parameter on the VNA and the VNA output power to EOM $P_{\textrm{EOM},\upmu}$, we calculate the output microwave flux from the device
\begin{equation}
\Ndotoutmu = \frac{P_{\textrm{EOM},\upmu}|S_{21}|^2}{\eta_{\textrm{cable}}\hbar \omegamu}.
\end{equation}

To calibrate the input optical sideband flux, we use a $1\%$ beamsplitter at the EOM output and a fiber Fabry-P\'erot tunable filter (FFP-TF, Micron Optics), and scan the FFP-TF to pickup the pump and sidebands separately. The filter output is detected by a Newport Nanosecond Photodetector, amplified and recorded.

\begin{figure}[htbp]
	\centering
	\includegraphics[scale=0.6]{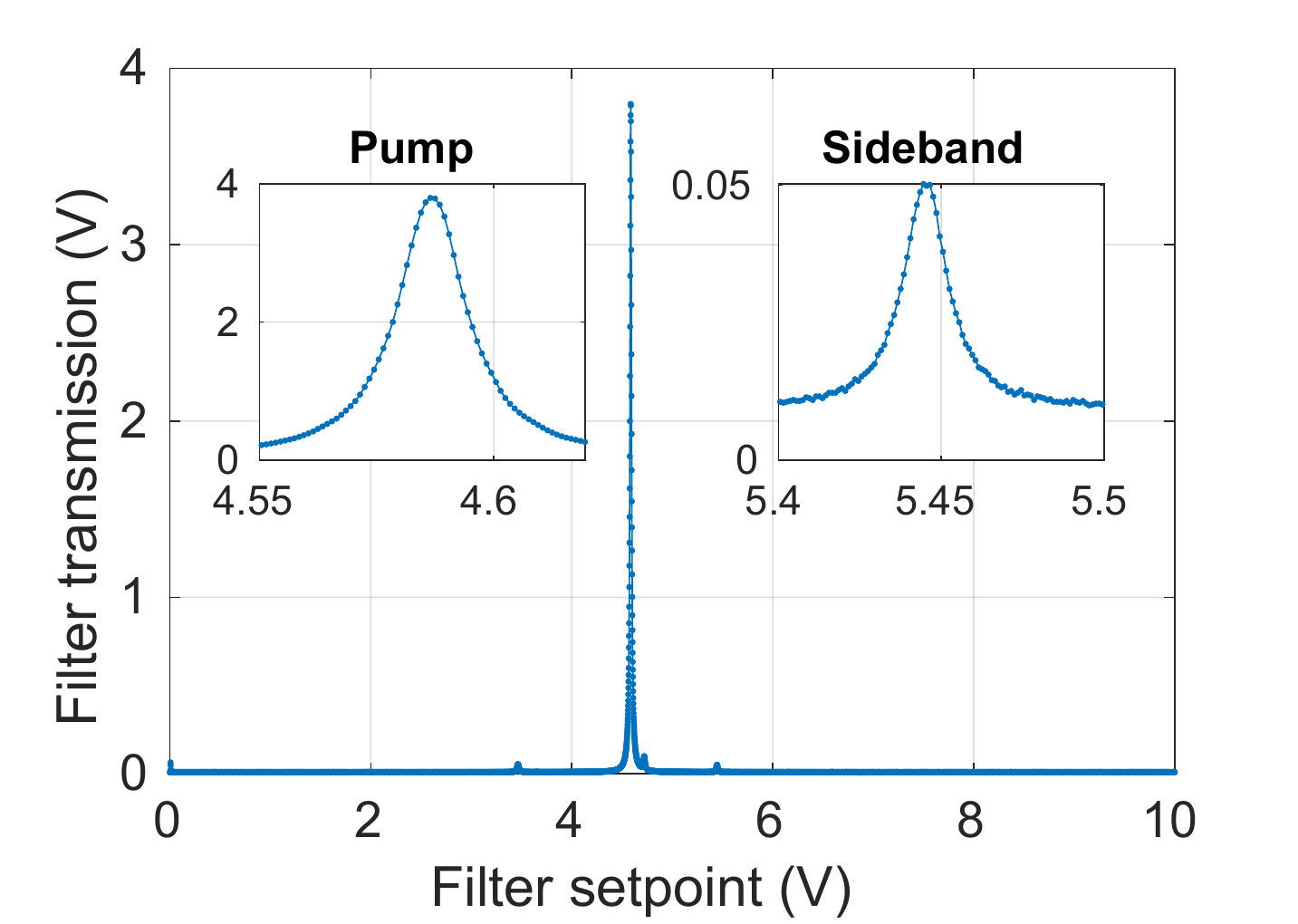}
	\caption{\label{SIFig:FilterScan} A typical filter scan. Left inset: zoomed-in plot of the pump. Right inset: zoomed-in plot of the lower frequency sideband.}
\end{figure}

A typical filter scan result is shown in Fig.~\ref{SIFig:FilterScan}. Higher filter setpoint voltage corresponds to longer filter pass wavelength. Two sidebands generated by the EOM are clearly visible with good signal-to-noise ratio. The small peak near the pump corresponds to a filter Fabry-P\'erot cavity mode with a different polarization and is nearly fully suppressed. The sideband ratio $ r\equiv P_{\textrm{sb}}/P_{\textrm{pump}} $ is measured from the ratio between the peak voltages of the relevant sideband and the pump, respectively. The dark voltage is subtracted before taking the ratio. Using $ 8~\textrm{dBm}$ VNA output microwave power to the EOM, we measure $r = (1.41\pm 0.08)\%$. The filter scan is taken for every different optical-to-microwave conversion measurement. Combining the sideband ratio and the measured input pump power $P_{\textrm{in,p}}$ before the lensed fiber, the input sideband photon flux is
\begin{equation}
\Ndotino = \frac{P_{\textrm{in,sb}}}{\hbar \omegac} = \frac{P_{\textrm{in,p}}r }{\hbar \omegac}.
\end{equation}

\section{Optical-to-microwave conversion measurement with low optical pump power}

\begin{figure*}[tb]
	\centering
	\includegraphics[scale=0.35]{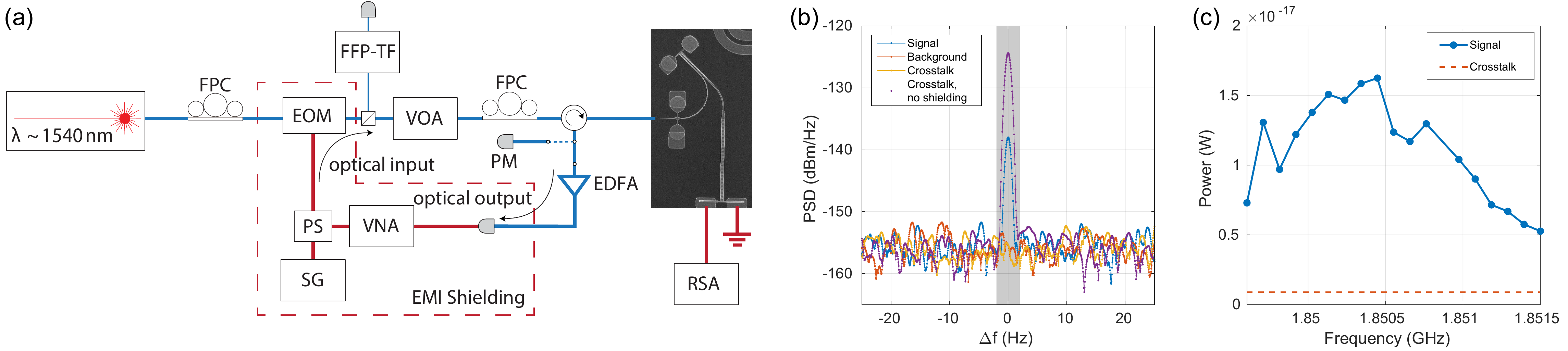}
	\caption{\label{SIFig:low-P-O2E} Low pump power optical-to-microwave conversion measurement. (a) Measurement setup. (b) A typical power spectral density (PSD) of the converted microwave signal. The RSA background and the crosstalk are also shown for comparison. The crosstalk is measured with SG on and zero input optical power. The EMI shielding reduced the crosstalk by more than $30$ dB. The shaded region shows the frequency range used for integrating the total microwave power. (c) Converted microwave power versus microwave frequency, from which a peak conversion efficiency $\etaeo = 1.09\times 10^{-5}$ is extracted.}
\end{figure*}

Due to a much lower energy per photon at the microwave frequency, the optical-to-microwave conversion is difficult to measure with VNA at low optical pump power. We adopt a different setup, using a microwave signal generator (SG) for the optical sideband generation and the RSA for detection of the converted microwave signal (Fig.~\ref{SIFig:low-P-O2E}(a)). The maximal output power of the SG is $P_{\textrm{max}} = 25~\textrm{dBm}$ and is much higher than the maximal output power of the VNA, bringing the optical sideband ratio $r$, defined as the ratio between the sideband optical power and the pump power, from $\sim 1\% $ to $\sim 8\%$. To access the optical sideband response for detuning locking, we keep the VNA-based optical intput and output measurement and use a power splitter (PS, Mini-Circuits ZFRSC-123-S+, insertion loss $\sim 10~\textrm{dB}$) to combine the signal from the VNA and the SG.

To detect the optical-to-microwave conversion with the maximal measured efficiency of $\sim 10^{-5}$ with red-detuned pump $P_{\textrm{pump}}\sim \SI{3}{\micro\watt}$, we need the microwave crosstalk level to be lower than the estimated converted microwave signal power $ P_{\upmu} \sim 2\times 10^{-17}~\SI{}{\watt} \sim -136~\textrm{dBm} $. Meanwhile, we measure an RSA internal noise floor level of $\sim -155~\textrm{dBm}/\SI{}{\hertz} $ and a microwave crosstalk $ P_{\textrm{c}}=3.8\times 10^{-16}~\SI{}{\watt} \sim -124~\textrm{dBm} $ with $25~\textrm{dBm}$ output from the SG to the EOM. The microwave crosstalk is determined to be mostly between the EOM and the on-chip electrodes and the RF probe that is in-touch with the electrodes. The crosstalk is not significant but is still much higher than the signal (Fig.~\ref{SIFig:low-P-O2E}(b)). To suppress the crosstalk, we utilize an electromagnetic interference (EMI) shielding that encloses all microwave instruments except the RSA. The EMI shielding brings the crosstalk level down to comparable or lower than the RSA noise floor (Fig.~\ref{SIFig:low-P-O2E}(b)).

We use a span of $\SI{50}{\hertz}$, a resolution bandwidth of $f_{\textrm{bw}} = \SI{1}{\hertz}$ and $10$ averages on the RSA, corresponding to a noise floor of $\sim -155~\textrm{dBm}/\SI{}{\hertz}  $ and a sweep time of $ \SI{50}{\sec}$ for a single trace of power spectrum. The SG frequency and RSA center frequency $f_{\upmu}$ are then swept simultaneously across $ \SI{2}{\mega\hertz}$ around the peak conversion frequency. The power spectrum of the converted microwave signal is recorded at every frequency. The cavity-laser detuning $ \Delta $ is verified and corrected before taking the next frequency point by the VNA optical sideband sweep. The total converted power is calculated by integrating the power spectral density within $ f_{\upmu} \pm 2 f_{\textrm{bw}}$ (Fig.~\ref{SIFig:low-P-O2E}(b)).

The peak conversion efficiency is calibrated similar to the method introduced in Sec.~\ref{sec:O2E-calib}. The optical sideband ratio $r = (7.5\pm 0.2)\%$ is measured by the FFP-TF, and the converted microwave photon flux is calculated from the integrated converted microwave power. We have taken into account the non-negligible power in the first blue and red sidebands and neglected higher order sidebands when calculating the sideband photon flux from the total optical power. $\nc$ is also calculated using the pump power when the modulation is active, which is $ \sim 13\% $ smaller than the total optical power.

\section{Thermally induced mechanical red shift}

\begin{figure}[htbp]
	\centering
	\includegraphics[scale=0.5]{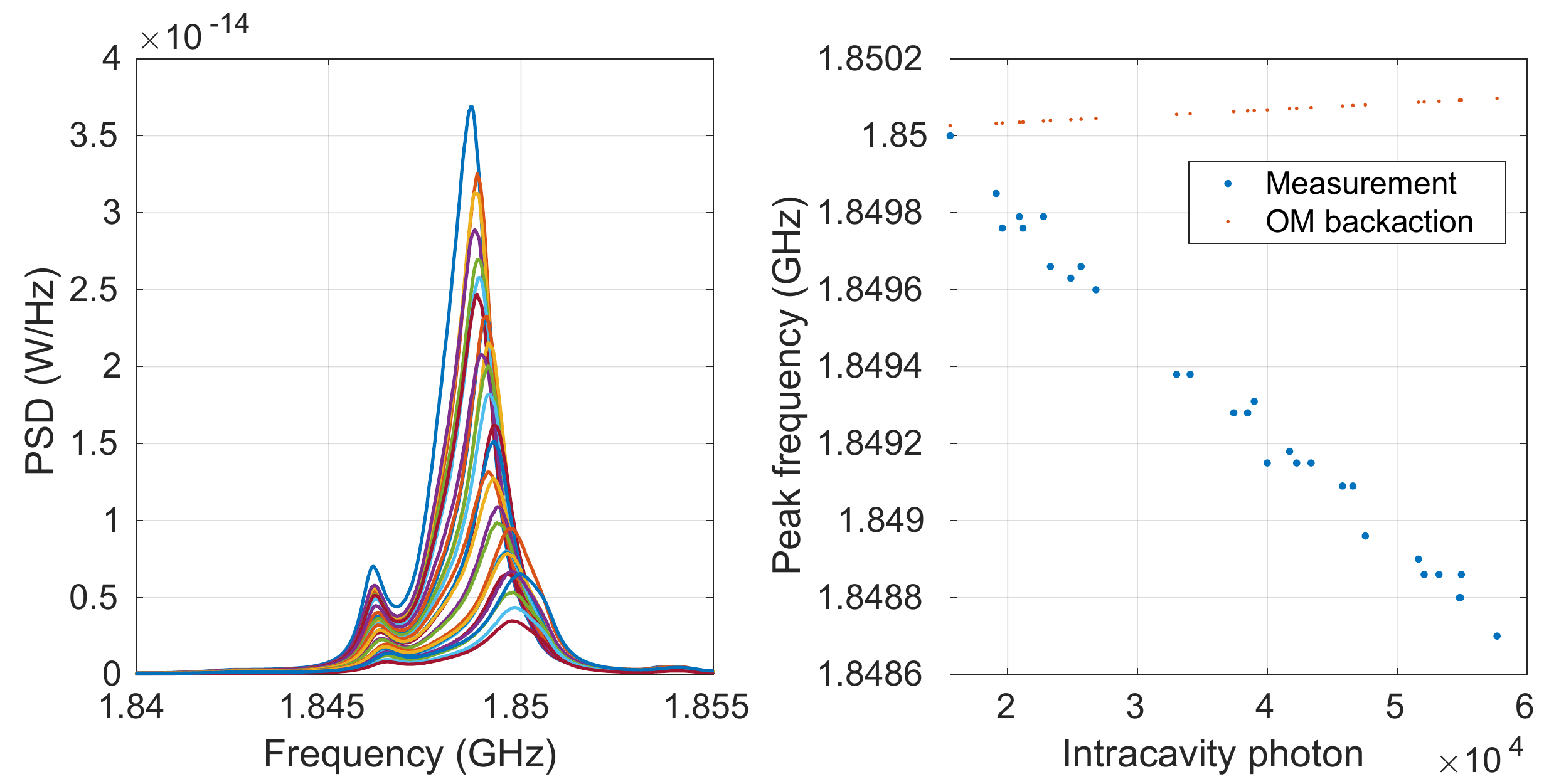}
	\caption{\label{SIFig:Thermal-mech-shift} Thermally induced mechanical frequency shift.}
\end{figure}

During the conversion measurement with the measurement setup shown in the main text, the thermal mechanical noise spectrum is recorded at the same time with the RSA. A clear red shift of the OMC mechanical mode is observed and is shown in Fig.~\ref{SIFig:Thermal-mech-shift}. Power spectral density (PSD) curves with different intracavity photon numbers at detuning $\Delta = -\omegam$ is plotted with different colors in linear scale. The extracted peak frequencies are shown on the right plot (blue). The expected optical stiffening from optomechanical backaction is also calculated for comparison (red) and is a minor effect comparing to the measured red shift. We attribute the red shift to thermal effect, where the thermal expansion caused by a higher temperature would increase device size and decrease the density, and lead to a lower mechanical mode frequency. The localized breathing mode and the mechanical waveguide modes have different overlaps with the heat profile generated by the optical mode and shift at different rates.

As a result, the OMC breathing mode not only slowly approaches the IDT center frequency, but also shifts faster and could go pass mechanical waveguide modes. We notice an asymmetric lineshape of the low power thermal mechanical PSD at low pump power as shown in Fig.~2(c) of the main text, where a waveguide mechanical mode frequency is almost overlapping with the OMC breathing mode but is at a slightly lower frequency. In Fig.~\ref{SIFig:Thermal-mech-shift}, the OMC breathing mode shifts across the same waveguide mode for increasing intracavity photon number from $\nc \sim 2\times 10^4 $ to $ \nc \sim 4\times 10^4$. The perfect overlap between the OMC breathing mode and a mechanical waveguide mode enhances the microwave-to-mechanical conversion efficiency and the total efficiency, which is clearly visible in Fig.~4(a) of the main text between $ 2\times 10^4 < \nc < 4\times 10^4$.

\end{document}